\documentclass[11pt,a4paper,english]{article}

\usepackage{a4}
\usepackage{babel}                     
\usepackage[latin1]{inputenc}
\usepackage{exscale}
\usepackage{amsmath}
\usepackage{amssymb}
\usepackage{graphicx}

\newlength{\defbaselineskip}
\newcommand{\ket}[1]{\ensuremath{\left| #1 \right\rangle}}
\newcommand{\bra}[1]{\ensuremath{\left\langle #1 \right|}}
\newcommand{\eval}[1]{\ensuremath{\left\langle #1 \right\rangle}}

\newcommand{\perm}[1]{
  P_{\!\!\scriptscriptstyle #1}
}

\newcommand{\fourpoints}{  
  \put(0,0){\circle*{1.5}}\put(5,0){\circle*{1.5}}\put(10,0){\circle*{1.5}}\put(15,0){\circle*{1.5}} 
}

\newcommand{\diagAone}{
\begin{picture}(19,7)
  \put(0,1.5){
  \begin{picture}(15,10)
  \fourpoints
  \put(0,0){\line(1,0){5}}
  \end{picture}}
\end{picture}
}

\newcommand{\diagAOneMod}{
\begin{picture}(19,7)
  \put(0,1.5){
  \begin{picture}(15,10)
  \fourpoints
  \put(10,0){\line(1,0){5}}
  \end{picture}}
\end{picture}
}

\newcommand{\diagAtwo}{
\begin{picture}(19,7)
  \put(0,1.5){
  \begin{picture}(15,10)
  \fourpoints
  \put(0,0){\line(1,0){5}}
  \put(10,0){\line(1,0){5}}
  \end{picture}}
\end{picture}
}

\newcommand{\diagBtwo}{
\begin{picture}(19,7)
  \put(0,1.5){
  \begin{picture}(15,10)
  \fourpoints
  \put(0,0){\line(1,0){5}}
  \put(5,0){\line(1,0){5}}
  \end{picture}}
\end{picture}
}

\newcommand{\diagCtwo}{
\begin{picture}(19,7)
  \put(0,1.5){
  \begin{picture}(15,10)
  \fourpoints
  \qbezier[18](0,0)(2.5,4)(5,0)
  \qbezier[18](0,0)(2.5,-4)(5,0)
  \end{picture}}
\end{picture}
}

\newcommand{\diagAthree}{
\begin{picture}(19,7)
  \put(0,1.5){
  \begin{picture}(15,10)
  \fourpoints
  \qbezier[18](0,0)(2.5,4)(5,0)
  \qbezier[18](0,0)(2.5,-4)(5,0)
  \put(5,0){\line(1,0){5}}
  \end{picture}}
\end{picture}
}

\newcommand{\diagBthree}{
\begin{picture}(19,7)
  \put(0,1.5){
  \begin{picture}(15,10)
  \fourpoints
  \put(0,0){\line(1,0){5}}
  \put(5,0){\line(1,0){5}}
  \qbezier[80](5,0)(10,7)(15,0)
  \end{picture}}
\end{picture}
}

\newcommand{\diagCthree}{
\begin{picture}(19,7)
  \put(0,1.5){
  \begin{picture}(15,10)
  \fourpoints
  \put(0,0){\line(1,0){5}}
  \put(5,0){\line(1,0){5}}
  \put(10,0){\line(1,0){5}}
  \end{picture}}
\end{picture}
}

\newcommand{\diagDthree}{
\begin{picture}(19,7)
  \put(0,1.5){
  \begin{picture}(15,10)
  \fourpoints
  \qbezier[18](0,0)(2.5,4)(5,0)
  \qbezier[18](0,0)(2.5,-4)(5,0)
  \put(0,0){\line(1,0){5}}
  \end{picture}}
\end{picture}
}

\newcommand{\diagEthree}{
\begin{picture}(19,7)
  \put(0,1.5){
  \begin{picture}(15,10)
  \fourpoints
  \qbezier[18](0,0)(2.5,4)(5,0)
  \qbezier[18](0,0)(2.5,-4)(5,0)
  \put(10,0){\line(1,0){5}}
  \end{picture}}
\end{picture}
}

\newcommand{\diagFthree}{
\begin{picture}(19,7)
  \put(0,1.5){
  \begin{picture}(15,10)
  \fourpoints
  \put(0,0){\line(1,0){5}}
  \put(5,0){\line(1,0){5}}
  \qbezier[36](0,0)(5,7)(10,0)
  \end{picture}}
\end{picture}
}

\newcommand{\diagAfour}{
\begin{picture}(19,7)
  \put(0,1.5){
  \begin{picture}(15,10)
  \fourpoints
  \put(5,0){\line(1,0){5}}
  \qbezier[36](0,0)(5,7)(10,0)
  \qbezier[36](0,0)(5,-7)(10,0)  
  \put(10,0){\line(1,0){5}}
  \end{picture}}
\end{picture}
}

\newcommand{\diagBfour}{
\begin{picture}(19,7)
  \put(0,1.5){
  \begin{picture}(15,5)
  \fourpoints
  \put(0,0){\line(1,0){5}}
  \put(10,0){\line(1,0){5}}
  \qbezier[18](5,0)(7.5,4)(10,0)
  \qbezier[18](5,0)(7.5,-4)(10,0)
  \end{picture}}
\end{picture}
}

\newcommand{\diagCfour}{
\begin{picture}(19,7)
  \put(0,1.5){
  \begin{picture}(15,10)
  \fourpoints
  \put(0,0){\line(1,0){5}}
  \put(5,0){\line(1,0){5}}
  \qbezier[36](0,0)(5,7)(10,0)
  \qbezier[36](0,0)(5,-7)(10,0)
  \end{picture}}
\end{picture}
}

\newcommand{\diagDfour}{
\begin{picture}(19,7)
  \put(0,1.5){
  \begin{picture}(15,10)
  \fourpoints
  \qbezier[18](0,0)(2.5,4)(5,0)
  \qbezier[18](0,0)(2.5,-4)(5,0)
  \put(5,0){\line(1,0){5}}
  \put(10,0){\line(1,0){5}}
  \end{picture}}
\end{picture}
}

\newcommand{\diagEfour}{
\begin{picture}(19,7)
  \put(0,1.5){
  \begin{picture}(15,10)
  \fourpoints
  \qbezier[18](0,0)(2.5,4)(5,0)
  \qbezier[18](0,0)(2.5,-4)(5,0)
  \qbezier[36](0,0)(2.5,7)(5,0)
  \qbezier[36](0,0)(2.5,-7)(5,0)
  \end{picture}}
\end{picture}
}

\newcommand{\diagFfour}{
\begin{picture}(19,7)
  \put(0,1.5){
  \begin{picture}(15,10)
  \fourpoints
  \qbezier[18](0,0)(2.5,4)(5,0)
  \qbezier[18](0,0)(2.5,-4)(5,0)
  \qbezier[18](5,0)(7.5,4)(10,0)
  \qbezier[18](5,0)(7.5,-4)(10,0)
  \end{picture}}
\end{picture}
}

\newcommand{\diagGfour}{
\begin{picture}(19,7)
  \put(0,1.5){
  \begin{picture}(15,10)
  \fourpoints
  \qbezier[18](0,0)(2.5,4)(5,0)
  \qbezier[18](0,0)(2.5,-4)(5,0)
  \put(0,0){\line(1,0){5}}
  \put(5,0){\line(1,0){5}}
  \end{picture}}
\end{picture}
}

\newcommand{\diagHfour}{
\begin{picture}(19,7)
  \put(0,1.5){
  \begin{picture}(15,10)
  \fourpoints
  \qbezier[18](0,0)(2.5,4)(5,0)
  \qbezier[18](0,0)(2.5,-4)(5,0)
  \put(0,0){\line(1,0){5}}
  \put(10,0){\line(1,0){5}}
  \end{picture}}
\end{picture}
}

\newcommand{\diagIfour}{
\begin{picture}(19,7)
  \put(0,1.5){
  \begin{picture}(15,10)
  \fourpoints
  \qbezier[18](0,0)(2.5,4)(5,0)
  \qbezier[18](0,0)(2.5,-4)(5,0)
  \qbezier[18](10,0)(12.5,4)(15,0)
  \qbezier[18](10,0)(12.5,-4)(15,0)
  \end{picture}}
\end{picture}
}

\newcommand{\diagJfour}{
\begin{picture}(19,7)
  \put(0,1.5){
  \begin{picture}(15,10)
  \fourpoints
  \put(0,0){\line(1,0){5}}
  \put(5,0){\line(1,0){5}}
  \put(10,0){\line(1,0){5}}
  \qbezier[36](0,0)(5,7)(10,0)
  \end{picture}}
\end{picture}
}

\newcommand{\diagKfour}{
\begin{picture}(19,7)
  \put(0,1.5){
  \begin{picture}(15,10)
  \fourpoints
  \put(0,0){\line(1,0){5}}
  \put(5,0){\line(1,0){5}}
  \put(10,0){\line(1,0){5}}
  \qbezier[36](0,0)(7.5,7)(15,0)
  \end{picture}}
\end{picture}
}

\newcommand{\diagAfive}{
\begin{picture}(19,7)
  \put(0,1.5){
  \begin{picture}(15,10)
  \fourpoints
  \qbezier[18](0,0)(2.5,4)(5,0)
  \qbezier[18](0,0)(2.5,-4)(5,0)
  \qbezier[18](5,0)(7.5,4)(10,0)
  \qbezier[18](5,0)(7.5,-4)(10,0)
  \qbezier[36](0,0)(5,7)(10,0)
  \end{picture}}
\end{picture}
}

\newcommand{\diagBfive}{
\begin{picture}(19,7)
  \put(0,1.5){
  \begin{picture}(15,5)
  \fourpoints
  \put(0,0){\line(1,0){5}}
  \put(10,0){\line(1,0){5}}
  \qbezier[18](5,0)(7.5,4)(10,0)
  \qbezier[18](5,0)(7.5,-4)(10,0)
  \put(5,0){\line(1,0){5}}
  \end{picture}}
\end{picture}
}

\newcommand{\diagCfive}{
\begin{picture}(19,7)
  \put(0,1.5){
  \begin{picture}(15,10)
  \fourpoints
  \qbezier[18](0,0)(2.5,4)(5,0)
  \qbezier[18](0,0)(2.5,-4)(5,0)
  \put(0,0){\line(1,0){5}}
  \put(5,0){\line(1,0){5}}
  \qbezier[36](0,0)(5,7)(10,0)  
  \end{picture}}
\end{picture}
}

\newcommand{\diagDfive}{
\begin{picture}(19,7)
  \put(0,1.5){
  \begin{picture}(15,10)
  \fourpoints
  \qbezier[18](0,0)(2.5,4)(5,0)
  \qbezier[18](0,0)(2.5,-4)(5,0)
  \put(0,0){\line(1,0){5}}
  \put(5,0){\line(1,0){5}}
  \put(10,0){\line(1,0){5}}  
  \end{picture}}
\end{picture}
}

\newcommand{\diagEfive}{
\begin{picture}(19,7)
  \put(0,1.5){
  \begin{picture}(15,10)
  \fourpoints
  \qbezier[18](0,0)(2.5,4)(5,0)
  \qbezier[18](0,0)(2.5,-4)(5,0)
  \qbezier[36](0,0)(2.5,7)(5,0)
  \qbezier[36](0,0)(2.5,-7)(5,0)
  \put(5,0){\line(1,0){5}}  
  \end{picture}}
\end{picture}
}

\newcommand{\diagFfive}{
\begin{picture}(19,7)
  \put(0,1.5){
  \begin{picture}(15,10)
  \fourpoints
  \qbezier[18](0,0)(2.5,4)(5,0)
  \qbezier[18](0,0)(2.5,-4)(5,0)
  \qbezier[18](5,0)(7.5,4)(10,0)
  \qbezier[18](5,0)(7.5,-4)(10,0)
  \put(0,0){\line(1,0){5}}
  \end{picture}}
\end{picture}
}

\newcommand{\diagGfive}{
\begin{picture}(19,7)
  \put(0,1.5){
  \begin{picture}(15,10)
  \fourpoints
  \qbezier[18](0,0)(2.5,4)(5,0)
  \qbezier[18](0,0)(2.5,-4)(5,0)
  \qbezier[18](5,0)(7.5,4)(10,0)
  \qbezier[18](5,0)(7.5,-4)(10,0)
  \put(10,0){\line(1,0){5}}
  \end{picture}}
\end{picture}
}

\newcommand{\diagHfive}{
\begin{picture}(19,7)
  \put(0,1.5){
  \begin{picture}(15,10)
  \fourpoints
  \qbezier[18](0,0)(2.5,4)(5,0)
  \qbezier[18](0,0)(2.5,-4)(5,0)
  \put(0,0){\line(1,0){5}}
  \qbezier[18](10,0)(12.5,4)(15,0)
  \qbezier[18](10,0)(12.5,-4)(15,0)
  \end{picture}}
\end{picture}
}

\newcommand{\diagIfive}{
\begin{picture}(19,7)
  \put(0,1.5){
  \begin{picture}(15,10)
  \fourpoints
  \qbezier[18](0,0)(2.5,4)(5,0)
  \qbezier[18](0,0)(2.5,-4)(5,0)
  \put(5,0){\line(1,0){5}}
  \qbezier[18](10,0)(12.5,4)(15,0)
  \qbezier[18](10,0)(12.5,-4)(15,0)
  \end{picture}}
\end{picture}
}

\newcommand{\diagJfive}{
\begin{picture}(19,7)
  \put(0,1.5){
  \begin{picture}(15,10)
  \fourpoints
  \put(0,0){\line(1,0){5}}
  \put(5,0){\line(1,0){5}}
  \qbezier[36](0,0)(5,7)(10,0)
  \qbezier[18](10,0)(12.5,4)(15,0)
  \qbezier[18](10,0)(12.5,-4)(15,0)
  \end{picture}}
\end{picture}
}

\newcommand{\diagKfive}{
\begin{picture}(19,7)
  \put(0,1.5){
  \begin{picture}(15,10)
  \fourpoints
  \put(0,0){\line(1,0){5}}
  \put(5,0){\line(1,0){5}}
  \qbezier[36](0,0)(5,7)(10,0)
  \qbezier[36](0,0)(5,-7)(10,0)
  \put(10,0){\line(1,0){5}}
  \end{picture}}
\end{picture}
}

\newcommand{\diagLfive}{
\begin{picture}(19,7)
  \put(0,1.5){
  \begin{picture}(15,10)
  \fourpoints
  \put(5,0){\line(1,0){5}}
  \qbezier[36](0,0)(5,7)(10,0)
  \qbezier[36](0,0)(5,-7)(10,0)  
  \qbezier[18](10,0)(12.5,4)(15,0)
  \qbezier[18](10,0)(12.5,-4)(15,0)
  \end{picture}}
\end{picture}
}

\newcommand{\diagMfive}{
\begin{picture}(19,7)
  \put(0,1.5){
  \begin{picture}(15,10)
  \fourpoints
  \qbezier[18](0,0)(2.5,4)(5,0)
  \qbezier[18](0,0)(2.5,-4)(5,0)
  \put(0,0){\line(1,0){5}}
  \put(5,0){\line(1,0){5}}
  \qbezier[36](5,0)(10,7)(15,0)  
  \end{picture}}
\end{picture}
}

\newcommand{\diagNfive}{
\begin{picture}(19,7)
  \put(0,1.5){
  \begin{picture}(15,10)
  \fourpoints
  \qbezier[18](0,0)(2.5,4)(5,0)
  \qbezier[18](0,0)(2.5,-4)(5,0)
  \qbezier[36](0,0)(2.5,7)(5,0)
  \qbezier[36](0,0)(2.5,-7)(5,0)
  \put(10,0){\line(1,0){5}}
  \end{picture}}
\end{picture}
}

\newcommand{\diagOfive}{
\begin{picture}(19,7)
  \put(0,1.5){
  \begin{picture}(15,10)
  \fourpoints
  \qbezier[36](0,0)(2.5,3)(5,0)
  \qbezier[36](0,0)(2.5,-3)(5,0)
  \put(5,0){\line(1,0){5}}
  \put(10,0){\line(1,0){5}}
  \qbezier[36](0,0)(7.5,7)(15,0)
  \end{picture}}
\end{picture}
}

\newcommand{\diagPfive}{
\begin{picture}(19,7)
  \put(0,1.5){
  \begin{picture}(15,10)
  \fourpoints
  \put(0,0){\line(1,0){5}}
  \qbezier[18](5,0)(7.5,4)(10,0)
  \qbezier[18](5,0)(7.5,-4)(10,0)
  \qbezier[36](0,0)(5,7)(10,0)
  \qbezier[36](0,0)(7.5,-7)(15,0)
  \end{picture}}
\end{picture}
}

\newcommand{\diagQfive}{
\begin{picture}(19,7)
  \put(0,1.5){
  \begin{picture}(15,10)
  \fourpoints
  \put(0,0){\line(1,0){5}}
  \put(5,0){\line(1,0){5}}
  \put(10,0){\line(1,0){5}}
  \qbezier[36](0,0)(5,7)(10,0)
  \qbezier[36](0,0)(7.5,-7)(15,0)
  \end{picture}}
\end{picture}
}

\newcommand{\diagAsix}{
\begin{picture}(19,7)
  \put(0,1.5){
  \begin{picture}(15,10)
  \fourpoints
  \qbezier[18](0,0)(2.5,4)(5,0)
  \qbezier[18](0,0)(2.5,-4)(5,0)
  \qbezier[18](5,0)(7.5,4)(10,0)
  \qbezier[18](5,0)(7.5,-4)(10,0)
  \qbezier[36](0,0)(5,7)(10,0)
  \qbezier[36](0,0)(5,-7)(10,0)  
  \end{picture}}
\end{picture}
}

\newcommand{\diagBsix}{
\begin{picture}(19,7)
  \put(0,1.5){
  \begin{picture}(15,10)
  \fourpoints
  \qbezier[18](0,0)(2.5,4)(5,0)
  \qbezier[18](0,0)(2.5,-4)(5,0)
  \qbezier[18](5,0)(7.5,4)(10,0)
  \qbezier[18](5,0)(7.5,-4)(10,0)
  \put(0,0){\line(1,0){5}}
  \qbezier[36](0,0)(5,7)(10,0)  
  \end{picture}}
\end{picture}
}

\newcommand{\diagCsix}{
\begin{picture}(19,7)
  \put(0,1.5){
  \begin{picture}(15,10)
  \fourpoints
  \qbezier[18](5,0)(7.5,4)(10,0)
  \qbezier[18](5,0)(7.5,-4)(10,0)
  \qbezier[36](5,0)(7.5,7)(10,0)
  \qbezier[36](5,0)(7.5,-7)(10,0)   
  \put(0,0){\line(1,0){5}}
  \put(10,0){\line(1,0){5}}  
  \end{picture}}
\end{picture}
}

\newcommand{\diagDsix}{
\begin{picture}(19,7)
  \put(0,1.5){
  \begin{picture}(15,10)
  \fourpoints
  \qbezier[18](0,0)(2.5,4)(5,0)
  \qbezier[18](0,0)(2.5,-4)(5,0)
  \qbezier[18](5,0)(7.5,4)(10,0)
  \qbezier[18](5,0)(7.5,-4)(10,0)
  \put(0,0){\line(1,0){5}}
  \qbezier[36](0,0)(7.5,9)(15,0)  
  \end{picture}}
\end{picture}
}

\newcommand{\diagEsix}{
\begin{picture}(19,7)
  \put(0,1.5){
  \begin{picture}(15,10)
  \fourpoints
  \qbezier[18](0,0)(2.5,4)(5,0)
  \qbezier[18](0,0)(2.5,-4)(5,0)
  \put(0,0){\line(1,0){5}}
  \qbezier[36](0,0)(2.5,-7)(5,0)
  \put(5,0){\line(1,0){5}}  
  \qbezier[36](0,0)(5,7)(10,0)  
  \end{picture}}
\end{picture}
}

\newcommand{\diagFsix}{
\begin{picture}(19,7)
  \put(0,1.5){
  \begin{picture}(15,10)
  \fourpoints
  \qbezier[18](0,0)(2.5,4)(5,0)
  \qbezier[18](0,0)(2.5,-4)(5,0)
  \qbezier[18](5,0)(7.5,4)(10,0)
  \qbezier[18](5,0)(7.5,-4)(10,0)
  \put(0,0){\line(1,0){5}}
  \put(10,0){\line(1,0){5}}  
  \end{picture}}
\end{picture}
}

\newcommand{\diagGsix}{
\begin{picture}(19,7)
  \put(0,1.5){
  \begin{picture}(15,10)
  \fourpoints
  \qbezier[18](0,0)(2.5,4)(5,0)
  \qbezier[18](0,0)(2.5,-4)(5,0)
  \qbezier[18](5,0)(7.5,4)(10,0)
  \qbezier[18](5,0)(7.5,-4)(10,0)
  \qbezier[18](10,0)(12.5,4)(15,0)
  \qbezier[18](10,0)(12.5,-4)(15,0)  
  \end{picture}}
\end{picture}
}

\newcommand{\diagHsix}{
\begin{picture}(19,7)
  \put(0,1.5){
  \begin{picture}(15,10)
  \fourpoints
  \qbezier[18](0,0)(2.5,4)(5,0)
  \qbezier[18](0,0)(2.5,-4)(5,0)
  \put(0,0){\line(1,0){5}}
  \put(5,0){\line(1,0){5}}
  \qbezier[18](10,0)(12.5,4)(15,0)
  \qbezier[18](10,0)(12.5,-4)(15,0)  
  \end{picture}}
\end{picture}
}

\newcommand{\diagIsix}{
\begin{picture}(19,7)
  \put(0,1.5){
  \begin{picture}(15,10)
  \fourpoints
  \qbezier[18](0,0)(2.5,4)(5,0)
  \qbezier[18](0,0)(2.5,-4)(5,0)
  \qbezier[36](0,0)(2.5,7)(5,0)
  \qbezier[36](0,0)(2.5,-7)(5,0)
  \qbezier[18](10,0)(12.5,4)(15,0)  
  \qbezier[18](10,0)(12.5,-4)(15,0)  
  \end{picture}}
\end{picture}
}

\newcommand{\diagJsix}{
\begin{picture}(19,7)
  \put(0,1.5){
  \begin{picture}(15,10)
  \fourpoints
  \qbezier[18](0,0)(2.5,4)(5,0)
  \qbezier[18](0,0)(2.5,-4)(5,0)
  \put(0,0){\line(1,0){5}}
  \put(10,0){\line(1,0){5}}
  \qbezier[18](10,0)(12.5,4)(15,0)
  \qbezier[18](10,0)(12.5,-4)(15,0)  
  \end{picture}}
\end{picture}
}

\newcommand{\diagKsix}{
\begin{picture}(19,7)
  \put(0,1.5){
  \begin{picture}(15,10)
  \fourpoints
  \put(5,0){\line(1,0){5}}
  \put(10,0){\line(1,0){5}}
  \qbezier[36](0,0)(5,-4)(10,0)
  \qbezier[36](0,0)(7.5,-7)(15,0)
  \qbezier[36](0,0)(7.5,7)(15,0)
  \qbezier[36](0,0)(7.5,4)(15,0)    
  \end{picture}}
\end{picture}
}

\newcommand{\diagLsix}{
\begin{picture}(19,7)
  \put(0,1.5){
  \begin{picture}(15,10)
  \fourpoints
  \put(0,0){\line(1,0){5}}  
  \put(5,0){\line(1,0){5}}
  \qbezier[36](0,0)(5,-4)(10,0)
  \qbezier[36](0,0)(7.5,-7)(15,0)
  \qbezier[36](0,0)(7.5,7)(15,0)
  \qbezier[36](0,0)(7.5,4)(15,0)    
  \end{picture}}
\end{picture}
}

\newcommand{\diagMsix}{
\begin{picture}(19,7)
  \put(0,1.5){
  \begin{picture}(15,10)
  \fourpoints
  \qbezier[36](0,0)(2.5,3)(5,0)
  \qbezier[36](0,0)(2.5,-3)(5,0)
  \put(5,0){\line(1,0){5}}
  \put(10,0){\line(1,0){5}}
  \qbezier[36](0,0)(5,7)(10,0)
  \qbezier[36](0,0)(5,-7)(10,0)  
  \end{picture}}
\end{picture}
}

\newcommand{\diagNsix}{
\begin{picture}(19,7)
  \put(0,1.5){
  \begin{picture}(15,10)
  \fourpoints
  \put(0,0){\line(1,0){5}}
  \put(5,0){\line(1,0){5}}
  \qbezier[36](0,0)(5,7)(10,0)
  \qbezier[18](10,0)(12.5,4)(15,0)
  \qbezier[18](10,0)(12.5,-4)(15,0)
  \qbezier[36](0,0)(7.5,-9)(15,0)  
  \end{picture}}
\end{picture}
}

\newcommand{\diagOsix}{
\begin{picture}(19,7)
  \put(0,1.5){
  \begin{picture}(15,10)
  \fourpoints
  \qbezier[18](0,0)(2.5,4)(5,0)
  \qbezier[18](0,0)(2.5,-4)(5,0)
  \qbezier[18](10,0)(12.5,4)(15,0)
  \qbezier[18](10,0)(12.5,-4)(15,0)
  \qbezier[36](0,0)(5,7)(10,0)  
  \qbezier[36](5,0)(10,-7)(15,0)    
  \end{picture}}
\end{picture}
}

\newcommand{\diagPsix}{
\begin{picture}(19,7)
  \put(0,1.5){
  \begin{picture}(15,10)
  \fourpoints
  \put(0,0){\line(1,0){5}}  
  \put(5,0){\line(1,0){5}}
  \put(10,0){\line(1,0){5}}
  \qbezier[36](0,0)(7.5,-7)(15,0)
  \qbezier[36](0,0)(7.5,7)(15,0)
  \qbezier[36](0,0)(7.5,4)(15,0)    
  \end{picture}}
\end{picture}
}

\newcommand{\diagQsix}{
\begin{picture}(19,7)
  \put(0,1.5){
  \begin{picture}(15,10)
  \fourpoints
  \qbezier[18](0,0)(2.5,4)(5,0)
  \qbezier[18](0,0)(2.5,-4)(5,0)
  \qbezier[18](5,0)(7.5,4)(10,0)
  \qbezier[18](5,0)(7.5,-4)(10,0)
  \put(10,0){\line(1,0){5}}
  \qbezier[36](5,0)(10,7)(15,0)  
  \end{picture}}
\end{picture}
}

\newcommand{\diagRsix}{
\begin{picture}(19,7)
  \put(0,1.5){
  \begin{picture}(15,10)
  \fourpoints
  \put(0,0){\line(1,0){5}}
  \put(5,0){\line(1,0){5}}
  \qbezier[36](0,0)(5,7)(10,0)
  \qbezier[36](0,0)(5,4)(10,0)  
  \qbezier[36](0,0)(5,-7)(10,0)
  \put(10,0){\line(1,0){5}}
  \end{picture}}
\end{picture}
}

\newcommand{\diagSsix}{
\begin{picture}(19,7)
  \put(0,1.5){
  \begin{picture}(15,10)
  \fourpoints
  \put(0,0){\line(1,0){5}}  
  \qbezier[18](5,0)(7.5,3)(10,0)
  \qbezier[18](5,0)(7.5,-3)(10,0)  
  \qbezier[36](5,0)(10,5)(15,0)      
  \qbezier[36](0,0)(7.5,-7)(15,0)
  \qbezier[36](0,0)(7.5,7)(15,0)
  \end{picture}}
\end{picture}
}

\newcommand{\diagTsix}{
\begin{picture}(19,7)
  \put(0,1.5){
  \begin{picture}(15,10)
  \fourpoints
  \qbezier[18](0,0)(2.5,4)(5,0)
  \qbezier[18](0,0)(2.5,-4)(5,0)
  \qbezier[18](5,0)(7.5,4)(10,0)
  \qbezier[18](5,0)(7.5,-4)(10,0)
  \qbezier[36](0,0)(5,7)(10,0)
  \qbezier[36](5,0)(10,-7)(15,0)  
  \end{picture}}
\end{picture}
}

\newcommand{\diagUsix}{
\begin{picture}(19,7)
  \put(0,1.5){
  \begin{picture}(15,10)
  \fourpoints
  \qbezier[18](0,0)(2.5,4)(5,0)
  \qbezier[18](0,0)(2.5,-4)(5,0)
  \qbezier[18](5,0)(7.5,4)(10,0)
  \qbezier[18](5,0)(7.5,-4)(10,0)
  \put(10,0){\line(1,0){5}}
  \qbezier[36](0,0)(7.5,7)(15,0)  
  \end{picture}}
\end{picture}
}

\newcommand{\diagVsix}{
\begin{picture}(19,7)
  \put(0,1.5){
  \begin{picture}(15,5)
  \fourpoints
  \put(0,0){\line(1,0){5}}
  \put(10,0){\line(1,0){5}}
  \qbezier[18](5,0)(7.5,4)(10,0)
  \qbezier[18](5,0)(7.5,-4)(10,0)
  \qbezier[36](0,0)(5,7)(10,0)  
  \qbezier[36](5,0)(10,-7)(15,0)    
  \end{picture}}
\end{picture}
}

\newcommand{\diagWsix}{
\begin{picture}(19,7)
  \put(0,1.5){
  \begin{picture}(15,10)
  \fourpoints
  \put(0,0){\line(1,0){5}}
  \put(5,0){\line(1,0){5}}
  \put(10,0){\line(1,0){5}}
  \qbezier[36](0,0)(7.5,7)(15,0)
  \qbezier[36](0,0)(5,4)(10,0)  
  \qbezier[36](5,0)(10,-4)(15,0)    
  \end{picture}}
\end{picture}
}

\begin{document}
\setlength{\unitlength}{0.7mm}

\begin{picture}(0,0)
	\put(170,20){hep-th/0504211}
\end{picture}

\begin{center}
  {\Large\textbf{Four-Point Functions in \\[0.5ex]
     Logarithmic Conformal Field Theories\\[1.3ex]}}
\end{center}

\vspace{1cm}

\begin{center}
\textsc{Michael Flohr$^{1,2}$ and Marco Krohn$^{2}$}

  \vspace{1cm}
  {\em $^1$ Physikalisches Institut, University of Bonn\\
       Nussallee 12, D-53115 Bonn, Germany }\\
       \texttt{flohr@th.physik.uni-bonn.de} \\
       
  \vspace{0.5cm}
  
  {\em $^2$ Institute for Theoretical Physics, University of Hannover\\
       Appelstraße 2, D-30167 Hannover, Germany} \\
       \texttt{krohn@itp.uni-hannover.de}
\end{center}

\vspace{1cm}

\begin{abstract}
The generic structure of 4-point functions of fields residing in
indecomposable representations of arbitrary rank is given. The used
algorithm is described and we present all results for Jordan-rank $r=2$ and
$r=3$ where we make use of permutation symmetry and use a graphical
representation for the results. A number of remaining degrees of freedom which
can show up in the correlator are discussed in detail. Finally we present
the results for two-logarithmic fields for arbitrary Jordan-rank.
\end{abstract}

\newpage
\tableofcontents

\newpage

\section{Introduction and formulation of the problem}
During the last few years, logarithmic conformal field
theory (LCFT) has been established as a well-defined variety of
conformal field theories in two dimensions. The defining feature of a LCFT
is the occurrence of indecomposable representations which, in turn, may lead to
logarithmically diverging correlation functions. 

The concept of LCFTs was considered in its own right first by Gurarie \cite{Gur93}.
Since then an enormous amount of work was done to understand LCFTs
and to link LCFTs to other fields in physics, see for example the 
reviews \cite{Flo01-11,Gab01} and references therein. The number 
of topics (logarithmic) conformal field theories might play a role 
in is still growing, e.\,g., there are suggestions
about links between Stochastic L\"owner evolutions (SLEs) 
and (L)CFTs. A nice review on SLE is \cite{Car05}, and a possible relation to
LCFT is discussed in \cite{Ras04}. 

Logarithmic conformal field theories are a generalization of 
conformal field theories (CFTs) in the sense that CFTs are LCFTs of 
Jordan-rank one. Since the works of Belavin, Polyakov and Zamolodchikov
in 1984 \cite{BPZ84} a powerful machinery of tools, algorithms and definitions
has been developed, which nowadays is indispensable for analyzing
conformal field theories. These definitions and techniques include 
characters, null vectors, operator product expansions (OPEs) and 
correlation functions, to name only a few. With the rise of LCFTs the demand 
for porting and generalizing these tools to LCFTs became an important
endeavor. Today, porting of definitions and techniques from CFTs to LCFTs
is almost finished, cf.\ \cite{Flo01, Nic02} and references therein.
Nevertheless there exist still some areas which are not well-understood, 
such as modular properties of characters and partition functions.

In the course of the paper we want to discuss the generic
form of four-point correlation functions, which is fixed by global conformal
invariance, in the case of LCFTs. The solution for this problem in case
of CFTs is well-known, but in case of LCFTs only incomplete results
exist so far. An example where four-point correlators play a role in, are
Abelian sandpile models which can be described by a $c=-2$
LCFT, e.\,g. \cite{Pir04, Mah01}.

In the case of ordinary conformal field theory (CFT) it is known that every 
correlation function has to 
fulfill the so called global conformal Ward identities (GCWI) as a 
consequence of invariance under global conformal transformations:
\begin{align}
  L_q \eval{\Psi_{(h_1)}(z_1) \ldots \Psi_{(h_n)}(z_n)}  & = 0, \quad q=-1,0,1 \; ,
\end{align}
where $\Psi_{(h)}(z)$ is a primary field and hence
\begin{align}
  L_q \eval{\Psi_{(h_1)}(z_1) \ldots \Psi_{(h_n)}(z_n)}  = &
    \sum_{i=1}^n z_i^q \left[ z_i\partial_i + (q+1)h_i \right] 
                                              \eval{ \ldots } \; .
\end{align}
When considering logarithmic conformal field theories, primary fields
appear together with so-called logarithmic partner fields which, in the
simplest case form indecomposable representations in the form of Jordan cells.
Then, these equations have to be slightly
altered, cf.\ \cite{Flo00}, by adding an additional term to the GCWI,
leading to the \emph{generalized global conformal Ward identities}:
\begin{equation}\label{gcwi_lcft}
  \sum_{i=1}^n z_i^q \left[ z_i\partial_i + (q+1)(h_i + \delta_{h_i}) \right] 
  \eval{\Psi_{(h_1, k_1)}(z_1) \ldots \Psi_{(h_n,k_n)}(z_n)} = 0 \;,
\end{equation}
where $\Psi_{(h_i,k_i)}(z_i)$ denotes a logarithmic field of Jordan-level $k_i$ respectively a primary
field in case $k_i = 0$. 
The operator $\delta_{h_i}$ acts on these logarithmic fields by reducing 
the Jordan-level of the field by $1$ respectively annihilating the field in case it is a primary one:
$\delta_{h_i} \Psi_{(h_i,k_i)} = \Psi_{(h_i,k_i-1)}$ for $k_i>1$ and $\delta_{h_i} \Psi_{(h_i,k_i=0)} = 0$
otherwise (field being a primary). Note that in the above equation the additional 
operator $\delta_{h_i}$ vanishes for $q=-1$ meaning that the LCFT version exactly matches 
the CFT version for this value of $q$.
The additional operator $\delta_{h_i}$
makes it much harder to find the generic form of the correlators,
because it renders the differential equations inhomogeneous, i.\,e.,
the solution will depend on solutions of lower Jordan-level. It is this
additional term $\delta_{h_i}$ that makes solving the equations a lot
harder compared to the CFT case.

If we consider the states corresponding to the fields $\Psi_{(h_i,k_i)}$, the 
action of $\delta_{h_i}$ leads to the following property for $L_0$ 
\begin{align}
  L_0 \ket{h;k} & = h \ket{h;k} + \ket{h;k-1} 
\end{align}
where additionally
\begin{align}  
  \ket{h;-k}  & = 0   \quad \forall k>0
\end{align}
holds. This shows, that the fields $\Psi_{(h_i,k_i)}$
indeed correspond to Jordan cells with respect to $L_0$.
The representation of a LCFT with the largest Jordan cell
defines the rank $r$ of the LCFT, i.\,e., $k_i<r$.

The representation space is, as usual, spanned by the 
states $\ket{h,k}$ defined by the field-state isomorphism
$\ket{h,k} := \lim_{z\rightarrow 0} \Psi_{(h,k)} \ket{0}$.
All these states are typically assumed to be quasi-primary in the sense
that $L_n\ket{h,k}=0\ \forall n>0$ and for all $k$. Thus, they almost
behave as highest-weight states, up to the non-diagonal action of $L_0$.
This is not true in general, because states to logarithmic partner fields 
may fail to be quasi-primary, i.\,e., $L_1\ket{h,k>0}\neq 0$. However, 
under certain assumptions, this does not affect the form of correlation 
functions. Furthermore, from the results for 1-, 2- and 3-point functions 
we can expect the vacuum representation to have 
the maximal Jordan-rank. No counter-examples are known
up to now and thus we assume that the Jordan-rank is the same
for all representations without loss of generality. The latter is justified
as follows: in case some smaller Jordan-rank representation does show
up, we can extend this representation by adding additional fields
which we set to zero. In essence, this simply means that the general
results remain valid with some of the structure constants set to zero.
For further details on the precise assumptions in the case of non
quasi-primary fields and on the maximal rank of the vacuum representation
see \cite{Flo01}.
 
While there are generic methods to determine 2- and 3-point correlation 
functions, e.\,g. see \cite{Flo01,GheKar97,KAR97,MRS00,RAK96}
and the particular elegant approach in \cite{Nag05}, no such method exists, to our 
knowledge, for 4-point correlation functions. However, in \cite{MRS00,MRS02} a
solution for the case of 4-point functions involving a level two null vector field
is given. 
On the other hand all $n>4$-point correlation functions can be reduced to 
2-, 3- and 4-point correlators. Therefore one can compute all observable 
quantities of a CFT--at least in principle--if one knows all 2-, 3- and 4-point 
functions. Thus, this work attempts to close the remaining gap by providing 
the prescription to fix the generic form of 4-point correlators in the case 
of arbitrary rank Jordan-cells in LCFT. 

While the generic form of 2- and 3-point functions is fixed up to structure 
constants 
the generic form of 4-point functions can be fixed only up to functions 
$F_{i_1 i_2 i_3 i_4}(x)$ of the
globally conformally invariant crossing ratios $x$. As in the case of ordinary 
conformal field theory these structure functions can be computed if additional
local symmetries, i.\,e., null vectors, exist. Indeed, such null vectors can exist
in the logarithmic case \cite{Flo97}, but the resulting differential equations are
more difficult to solve because they are inhomogeneous in general \cite{Flo00}.

In this paper we describe how the most general ansatz can be constructed and how
the emerging constants can be calculated in order to find a valid ansatz for equation (\ref{gcwi_lcft}).
Most of the constants can be fixed with the help of the global conformal Ward identities,
but we will also encounter cases where some degrees of freedom are left. A necessary
condition for these additional degrees of freedom is that all four fields in the four-point function
are of logarithmic origin. The number of degrees of freedom very much depends on the
form of the correlator. Furthermore we find that we have to identify some of 
the structure functions $F_{i_1 i_2 i_3 i_4}$ that are part of the correlator.

We then will use the discussed methods to determine all correlators for 
Jordan-rank $r=2$ and $r=3$. The results are given in a graphical representation
and also we make use of permutation symmetries in order to keep the
terms as short as possible. In the last section we consider the special
case that only two of the four fields are logarithmic and we show how
the resulting equations can be solved in this case for arbitrary Jordan-rank $r$.

\section{Approaching the problem}
In this section we describe how we simplify the initial problem 
and what algorithm we use to solve it for a Jordan-rank $r=2$ and $r=3$ theory.
We also discuss the appearance of additional degrees of freedom
that may show up if all four fields are of logarithmic type. For understanding
of this section it might be helpful to have a glance at the next section
which in detail discusses the most simple non-trivial case, that is Jordan-rank $r=2$.

\subsection{Simplification}
As motivated in the introductory section it is sufficient to consider
four-point functions and this is what we will do in the following.
We also noted in the introduction that equation (\ref{gcwi_lcft}) is equal to
the global conformal Ward identity in CFT for $q=-1$, meaning that the ansatz 
has to be translation invariant. Thus the ansatz depends on$z_{ij} := z_i - z_j$.
Thus the ansatz has the following form
\begin{align}
  \eval{\Psi_{(h_1,k_1)}(z_1) \ldots \Psi_{(h_4,k_4)}(z_4)} & = \prod_{i<j} z_{ij}^{\mu_{ij}} \, f( z_{12}, z_{13}, z_{14}, z_{23}, z_{24}, z_{34} ) \; ,
\end{align}
where the exponents $\mu_{ij} = \mu_{ji}$ have to satisfy the conditions
\begin{align}
  \sum_{j \ne i} \mu_{ij} = -2 h_i \; .
\end{align}
The factor $\prod_{i<j} z_{ij}^{\mu_{i}}$ exists to counter the $h_i$ terms 
on the left hand side of equation (\ref{gcwi_lcft}) and therefore we can 
without loss of generality set all conformal weights to zero, $h_i = 0$.
Note that the full correlator of course depends on the conformal weights. The
point here is that the global symmetries are not sufficient to fix the complete correlator,
but they are strong enough to fix the generic form and this form has no dependence on $h_i$. 
Therefore, we can simplify the resulting formulas by 
omitting the trivial direct dependency on the conformal weights.
If we set all conformal weights to zero then (\ref{gcwi_lcft}) becomes
\begin{equation}\label{gcwi_lcft_h=0}
  \sum_{i=1}^n z_i^q \left[ z_i\partial_i + (q+1) \delta_{h_i} \right]   \eval{k_1 k_2 k_3 k_4} = 0 \; ,
\end{equation}
where we write $k_i$ instead of the much longer form $\Psi_{(h_i,k_i)}(z_i)$.
The remaining two equations for $q=0,1$ have a $\delta_{h_i}$ term acting on the correlator
and thus lowering the sum of the Jordan-levels by one. Because of calculating
the expressions recursively we can assume the predecessors $\delta_{h_i} \eval{\ldots}$ to be known.
This leads to the final form
\begin{align}
  O_0 \eval{\dots} := \sum_{i=1}^4 z_i \partial_i \eval{\ldots} & = - \sum_i \delta_{h_i} \eval{\ldots}  \label{gcwi_lcft0} \\
  O_1 \eval{\dots} := \sum_{i=1}^4 z_i^2 \partial_i \eval{\ldots} & = -2 \sum_i z_i \delta_{h_i} \eval{\ldots} \; , \label{gcwi_lcft1}
\end{align}
where the correlators depend on the difference $z_{ij}$ only.
Though looking simple for given predecessors $\delta_{h_i} \eval{\ldots}$ at first glance, it is not easy
to find an ansatz for the correlator at all. Moreover we will learn that in some cases
the result is not unique. We sometimes use the sloppy term ``integrating" the predecessors $\delta_{h_i} \eval{\ldots}$
as a shortage for finding an ansatz that fulfills the above equations.

The starting point for the recursion is given by 
\begin{align}
  \eval{k_1 k_2 k_3 k_4} = F_0(x) \quad & \mathrm{for} \quad \sum_i k_i = r-1 \quad \mathrm{respectively} \label{eq:initcond1}\\
  \eval{k_1 k_2 k_3 k_4} = 0 \quad         & \mathrm{for} \quad \sum_i k_i < r-1\label{eq:initcond2}
\end{align}
where $x$ is the anharmonic ratio of the four points, $x = \tfrac{z_{12} z_{34}}{z_{14}z_{32}}$.
In essence this means that a correlation function with \emph{total Jordan-level} $K:=\sum_i k_i = r-1$
behaves like a correlation function in ordinary conformal field theory, i.\,e., 
it depends on one function of the globally conformally invariant anharmonic ratio.

The reason for these initial conditions comes from the fact that the only non-vanishing
one-point function in LCFT is the one of the highest level logarithmic partner of the
identity, $\Psi_{(h=0,k=r-1)}$. Evaluating a correlation function amounts to contracting
the inserted fields, in all possible ways, down to a one-point function.
Therefore, it is only natural to expect that the total Jordan-level $K$ of a 
non-vanishing correlator must at least be equal to $r-1$. Furthermore, since the cluster
decomposition property should hold, the initial conditions must also hold
for arbitrarily factorized correlators, e.\,g., $\eval{k_1 k_2 k_3 k_4} \sim \eval{k_1 k_2|0} \! \eval{0|k_3 k_4}$
in case that $z_1, z_2$ are well separated from $z_3, z_4$.
However, some care has to be taken about the correct insertion of the 
``identity" channel, which formally can be thought of to be of the form 
$\ket{0}\!\bra{0} = \sum_{k=0}^{r-1} \ket{h=0;k}\! \bra{h=0;r-1-k}$.
It is easy to see that the cluster decomposition with the above identity channel
implies (\ref{eq:initcond2}) and that precisely one term of this identity channel survives
yielding (\ref{eq:initcond1}), where we made use of the results for two-point functions
in \cite{Flo01}.

In the beginning we mentioned that $k_i > 0$ represents a logarithmic partner field,
while $k_i = 0$ is a primary field. We can subdivide the class of primary fields into two subclasses, the
so called proper primary-fields and the pre-logarithmic fields. This difference between
the subclasses becomes apparent if one considers the operator product expansion (OPE).
In contrast to the OPE of two proper primary-fields the OPE of two pre-logarithmic shows 
an additional term of logarithmic behavior, cf.\ \cite{KogLew97}.

In the following we consider proper primary-fields only 
and use the term synonymous with primary field. Restricting to
proper primary-fields is for simplicity only. It is possible to 
include pre-logarithmic fields into the theory, by making changes to
the initial condition (\ref{eq:initcond1}), (\ref{eq:initcond2}). For
instance in the well-known $c=-2$ example the initial-conditions for 
Jordan-rank $r=2$ would be
\begin{align}
  \eval{\phi\phi\phi\phi} & = 0 \; , \\
  \eval{ \mu \phi \phi \phi } & = 0 \; , \\
  \eval{ \mu \mu \phi \phi } &= F_0(x) \; ,
\end{align}
where $\phi$ stands for a proper primary and $\mu$ denotes a twist field.
Note that the same could be formally achieved by assigning
rational values $k_i$ to pre-logarithmic values, e.\,g., in this example
assigning a value of $k_i = \tfrac 12$ to the twist fields 
and using (\ref{eq:initcond1}), (\ref{eq:initcond2}) would lead to the
same initial conditions. A more precise analysis of this
and how to assign correct values for the $k_i$ can be found in \cite{Flo01}.
Apart from the initial conditions we also need slight adaption
of the ``connection rules" we are going to explain in subsection
\ref{sec:algorithm}. More comments can be found in the conclusions.

\subsection{Naming conventions}
The dependence of $F$ on the anharmonic ratio, is suppressed in the following. 
Further note that we do not write out the dependence on 
the Jordan-rank $r$, e.\,g., $\eval{1000} = F_0$ (for $r=2$)
as well as $\eval{1100} = \ldots = \eval{2000} = F_0$, namely for $r=3$.

As we will see the solution for all other total Jordan-levels $K := \sum_i k_i > r-1$ 
is always of the form
\begin{align}
  \eval{k_1 k_2 k_3 k_4} = & F_{k_1 k_2 k_3 k_4} + ( c_1 l_{12} + \ldots + c_6 l_{34} ) F_{k_1-1,k_2,k_3,k_4} + \ldots + \nonumber\\ 
 & (\mbox{logarithmic degree $K-r+1$}) F_0 \; ,
\end{align}
where $l_{ij} := \log(z_{ij})$. The highest logarithmic powers that appear in the solution are
always the factors associated with the function $F_0$. The degree in $l_{ij}$ also called
\emph{logarithmic degree} for short, is given by
\begin{align}\label{eq:deg log}
  \deg( l_{12}^{a_1} l_{13}^{a_2} \ldots l_{34}^{a_6} ) = \sum_i a_i \le K - r +1 =: l^{\mathrm{max}} \; .
\end{align}
There are cases where we will find
that some of the functions $F_{j_1 j_2 j_3 j_4}$ can be identified with each other, e.\,g., we will 
find that $F_{2100} \equiv F_{1200}$ for $r=3$. After identification we will always use 
the $F$-term whose index represents the lowest ``number". For example we write
$\eval{2100} = F_{1200}(x) + \ldots$ instead of using \mbox{$F_{2100}(x)$}.

In many places we decided to use a graphical representation instead of writing long
expressions of logarithms. The idea for this stems from \cite{Flo01-11} where it was 
chosen in order to give a better understanding of the contractions that can appear. 
Reading the diagrams is straightforward, the points stand for the four vertices and each
$l_{ij}$ is represented by a line between the vertices $i$ and $j$. Permutation operators $P$
are used to further reduce the length of the expressions, for instance
\begin{align}
  l_{12}^2 l_{23} l_{34} - l_{23}^2 l_{12} l_{14} = (1-P_{(13)}) \diagDfour \; .
\end{align}
From section \ref{sec:results for r=2} on we will always
use the graphical representation to present the results.

\subsection[Properties of $O_0$, $O_1$]{Properties of $\boldsymbol{O}_{\boldsymbol{0}}$, $\boldsymbol{O}_{\boldsymbol{1}}$}\label{properties of O}
Both operators $O_q$ are linear, nilpotent, act as derivatives on the function space and are
invariant under any permutations $p \in S_4$. 
The function space we consider is the space of polynomials in the logarithmic 
functions $l_{ij} := \log|z_i-z_j|$, called
$\mathcal{F}_{\mathrm{log}} := \mathbb{C}[l_{12},l_{13},l_{14},l_{23},l_{24},l_{34}]$.

For $q=0,1$ the operators $O_q$ have a simple behavior, when acting
on $\mathcal{F}_{\mathrm{log}}$:
\begin{align}
  O_0 : & \left\{  \begin{array}{ll}
    \mathcal{F}_{\mathrm{log}} \rightarrow \mathcal{F}_{\mathrm{log}} \\
    l_{i_1 j_1} \ldots l_{i_n j_n} \mapsto  \sum\limits_{k=1}^n l_{i_1 j_1} \ldots l_{i_{k-1} j_{k-1}} \, l_{i_{k+1} j_{k+1}} \ldots l_{i_n j_n} 
  \end{array}\right. \; , \label{eq:action of O0}\\
  O_1 : & \left\{  \begin{array}{ll}
    \mathcal{F}_{\mathrm{log}} \rightarrow \mathcal{F}_{\mathrm{log}}[\{z_{ij}\}] \\
    l_{i_1 j_1} \ldots l_{i_n j_n} \mapsto  \sum\limits_{k=1}^n l_{i_1 j_1} \ldots l_{i_{k-1} j_{k-1}} (z_{i_k} \!+\! z_{j_k}) l_{i_{k+1} j_{k+1}} \ldots l_{i_n j_n} 
  \end{array}\right. \; ,\label{eq:action of O1}
\end{align}
meaning that we can replace the term by a sum, where each $l_{ij}$ is replaced by either $1$ (for $q=0$)
or by $z_i + z_j$ (for $q=1$). Thus acting with $O_q$ on any term obviously reduces the logarithmic degree
by one and by that proves (\ref{eq:deg log}). 

An obvious question is whether the map $O_q : f \rightarrow f'$ is injective:
are there any non-trivial $f \in \mathcal{F}_{\mathrm{log}}$  with $O_0 f = 0$ and $O_1 f = 0$?
If we restrict ourselves to the function space $\mathcal{F}_{\mathrm{log}}$ then we find
that we can exactly determine the kernel of the operator $O := ( O_0, O_1 )$.

As will be shown in subsection 2.6 below, the kernel is given as follows.
\begin{align}
  \ker_{\mathcal{F}_{\mathrm{log}, g}} O & = \left\{ \sum_{i=0}^g a_i K_1^i K_2^{g-i} : a_k \in \mathbb{R} \right\} \; , \label{eq:form of kernel} \\
  K_1 & := l_{12} + l_{34} - l_{13} - l_{24} \label{eq:K1} \; , \\
  K_2 & := l_{12} + l_{34} - l_{14} - l_{23} \label{eq:K2} \; ,
\end{align}
where $\mathcal{F}_{\mathrm{log},g} := \{ f \in \mathcal{F}_{\mathrm{log}} | \deg f = g \}$
denotes the space of functions with logarithmic degree $g$, such that 
$\mathcal{F}_{\mathrm{log}} = \bigcup_g \mathcal{F}_{\mathrm{log},g}$.

\subsection{An ansatz for the equations}
As mentioned before we want to recursively solve the equations (\ref{gcwi_lcft0}) and (\ref{gcwi_lcft1}).
Since the number of terms quickly becomes huge and calculation tedious we make
use of computer algebra software for performing the calculations. 
In the next subsection we explain in more detail what we mean by
recursion. After this we show that the two equations can be reduced 
to a set of simpler equations and in subsection \ref{sec:algorithm} we 
present the algorithm we used for creating an ansatz.

\subsubsection{Recursion}\label{sec:recursion}
With recursion we mean the following: we start with the initial conditions 
as given in (\ref{eq:initcond1}) which corresponds to logarithmic degree $l=0$.
Then we calculate all necessary correlators which contain exactly one more
logarithmic field or one field whose Jordan-level is increased exactly by one.
In short this means that we determine all correlators of logarithmic degree $l=1$. 
The following diagram describes which correlators need to be calculated in 
order to determine the correlation function for $\eval{2110}$.
\begin{center}
\begin{picture}(143,60)
  \put(0,55){\eval{0110} = \eval{1100} = \eval{1010} = \eval{2000} = $F_0$ \qquad (l=0)} 
  \put(9,53){\vector(1,-1){20}}
  \put(31,53){\vector(0,-1){20}}
  \put(53,53){\vector(-1,-1){20}}  
  \put(33,53){\vector(1,-1){20}}
  \put(55,53){\vector(1,-1){20}}  
  \put(74,53){\vector(-1,-1){20}}
  \put(77,53){\vector(0,-1){20}}
  
  \put(0,27){\phantom{\eval{0110}} \phantom{=} \eval{1110} \phantom{=} \eval{2100} $\sim$ \eval{2010} \phantom{= $F_0$ }\qquad (l=1) }  
  \put(33,25){\vector(1,-1){20}}
  \put(55,25){\vector(0,-1){20}} 
  \put(77,25){\vector(-1,-1){20}} 
  \put(48,-1){\eval{2110}}
  \put(112,-1){(l=2)}
\end{picture}
\end{center} 
The effort for calculation can be reduced, since many of the correlators are
related $\sim$ to others by simple permutations, e.\,g., $\eval{2100} = \perm{23} \eval{2010}$.

\subsubsection{Breaking down into a set of equations}\label{sec:set of equations}
The operators $O_0$, $O_1$ in equations (\ref{gcwi_lcft0}), (\ref{gcwi_lcft1}) are linear, 
they act as derivatives on the correlators $\eval{\ldots}$ and they are invariant under 
any permutation $P \in S_4$ of the indices. The ansatz as well as the term on
the right hand side can, as we have seen before, be written in terms of the functions of $F_{\ldots}$, resulting in
\begin{eqnarray}
  O_q \big\{ F_{k_1 k_2 k_3 k_4} +  (\ldots)^u F_{k_1-1,k_2,k_3,k_4} +  (\ldots)^u F_{k_1,k_2-1,k_3,k_4} + \ldots \nonumber \\
     \ldots + (\ldots)^u F_{r-1,0,1,0} + (\ldots)^u F_{r-1,0,0,1} + (\ldots)^u F_0 \big\} & = & \nonumber \\
     (\ldots) F_{k_1-1,k_2,k_3,k_4} +  (\ldots) F_{k_1,k_2-1,k_3,k_4} + \ldots + (\ldots) F_{r-1,0,1,0} & & \nonumber \\
      + (\ldots) F_{r-1,0,0,1} + (\ldots) F_0 \; . \label{gcwi_lcft_expanded}
\end{eqnarray}
The terms $(\ldots)$ denote functions which may additionally depend on the differences $z_{12}, z_{13}, \ldots, z_{34}$
caused by the action of $O_1$. As usual $r$ is the Jordan-rank of the theory. For the right hand side we can assume 
these terms to be known, because we will solve the equations recursively. The corresponding 
terms on the left hand side are unknown, they are marked with a small ``$u$".
$O_q$ operates as a derivative and since $O_q F = 0$, we find that the problem reduces 
to ``integrating" the following set of equations
\begin{align}
  O_q (\ldots)^u_{k_1,k_2,k_3,k_4} & = 0 \; , \label{eq:gcwi set of eqns leading} \\
  O_q (\ldots)^u_{k_1-1,k_2,k_3,k_4} & = (\ldots)_{k_1-1,k_2,k_3,k_4} \; , \nonumber \\
  O_q (\ldots)^u_{k_1,k_2-1,k_3,k_4} & = (\ldots)_{k_1,k_2-1,k_3,k_4} \; ,\nonumber \\
  & \ldots \nonumber \\
  O_q (\ldots)^u_{r-1,0,0,1} & = (\ldots)_{r-1,0,0,1} \; , \nonumber \\
  O_q (\ldots)^u_{0} & = (\ldots)_{0} \; .\label{eq:gcwi set of eqns}
\end{align}
The upper index $u$ just reminds us that these terms are not yet known, and 
the lower index tells us from which part of the equation (\ref{gcwi_lcft_expanded}) the term stems from. 
Note that the first equation (\ref{eq:gcwi set of eqns leading}) and its solution
is well known
\begin{align}
  (\ldots)^u_{k_1,k_2,k_3,k_4} = F_{k_1,k_2,k_3,k_4}(x) \; ,
\end{align}
with $x$ being the anharmonic ratio.

\subsubsection{Description of the algorithm}\label{sec:algorithm}
Until now we did not specify what ansatz we fill in the left hand side
of the equations (\ref{eq:gcwi set of eqns leading}) to (\ref{eq:gcwi set of eqns}).
From OPE considerations \cite{Flo01} respectively from the structure
of the operators $O_0$ and $O_1$ we expect the correlators to consist of
terms of the type $l_{12}^{a_1} l_{13}^{a_2} \ldots l_{34}^{a_6}$, where each
term comes with an coefficient which needs to be determined. More precisely,
the generic structure of 2- and 3-point functions depends on the $l_{ij}$ in
a strictly polynomial form in such a way that the same is true for the
operator product expansion. Thus, also the 4-point functions should depend
only in a polynomial way on the $l_{ij}$ since, asymptotically, a 4-point 
function decomposes into an operator product expansion times remaining 3-point
functions, all of which are entirely polynomial in the $l_{ij}$. 
Unfortunately, the number of possible monomials in the $l_{ij}$ grows
heavily with the rank $r$ of the LCFT, and thus the number of 
coefficients.  
Luckily we can reduce the number of possible terms that can show up
in the following.

We do not have to take into account every logarithmic degree $a_1+a_2+\ldots+a_6$.
The equations (\ref{eq:action of O0}) and (\ref{eq:action of O1}) 
tell us that the logarithmic degree is reduced by one if we apply $O_0$ or $O_1$. 
If we assume for a moment that in every equation of (\ref{eq:gcwi set of eqns leading}) to (\ref{eq:gcwi set of eqns})
the right hand side consists of terms of the same logarithmic degree $l$, then it is 
apparent that the terms on the left hand side have logarithmic degree $l+1$.
We build all correlators recursively as explained in subsection \ref{sec:recursion}
and since our initial conditions only consists of one term on the right hand side,
we trivially find our assumption fulfilled. Thus by induction all
terms $l_{12}^{a_1} l_{13}^{a_2} \ldots l_{34}^{a_6}$ 
in  $(\ldots)^u_{n_1,n_2,n_3,n_4}$ have to have the same
logarithmic degree. 

As described in \cite{Flo01-11} it is helpful to use a graphical representation 
where each field $\Psi_{(h,k)}(z)$ in a Jordan-cell is depicted by a vertex
with $k$ outgoing lines. Contractions of logarithmic fields give rise to logarithms
in the correlators, where the possible powers with which $l_{ij}$ may
occur are determined by graph combinations. 
\begin{center}
  \includegraphics[width=6cm]{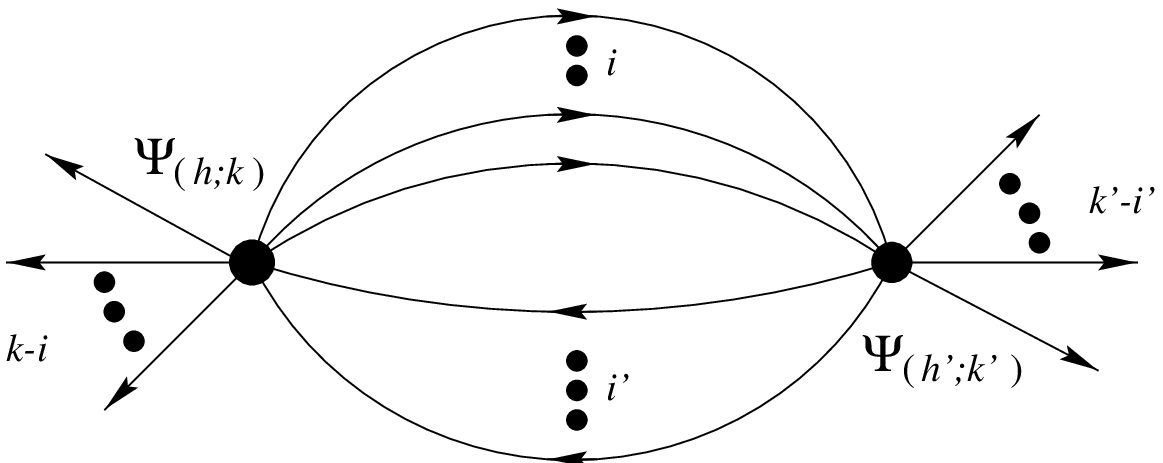}
\end{center}
Essentially, the terms for an ansatz of logarithmic degree $l$
are given by a sum over all admissible graphs subject to the 
following rules:
\begin{enumerate}
  \item use at most $k_i$ legs of a vertex for connections with other vertices,
  \item the source $i$ and the destination vertex $j$ have to be different: $i \ne j$,
  \item do connections with other logarithmic fields only (do not connect with primary fields) ,
  \item create exactly $l$ connections ,
  \item write $l_{ij}$ for every connection between two vertices $i$ and $j$. 
\end{enumerate}
Let us have a look at a simple example. We consider a theory of Jordan-rank $r=3$
and are interested in the structure of the correlator $\eval{2110}$ for the highest possible logarithmic
degree, i.\,e., the $F_0$ term. The corresponding graph for $\eval{2110}$ is
\begin{center}
\begin{picture}(19,7)
  \put(0,1.5){
  \begin{picture}(15,10)
  \fourpoints
  \put(0,0){\line(-1,3){2}}
  \put(0,0){\line(1,3){2}}
  \put(5,0){\line(1,3){2}}
  \put(10,0){\line(1,3){2}}
  \put(22,0){.}
  \end{picture}}
\end{picture} 
\end{center} 
Altogether we have four legs to our disposal, but we
also have to fix two of them leaving us with two free legs.
If we want to know which terms can appear for logarithmic degree $l=2$,
then we have to create all 2-contractions according to the above rules.
This results in the following six different graphs:
\begin{center}
\begin{picture}(19,7)
  \put(0,1.5){
  \begin{picture}(15,10)
  \fourpoints
  \put(0,0){\line(1,0){5}}
  \qbezier[80](0,0)(5,7)(10,0)
  \end{picture}}
\end{picture}
, \;
\begin{picture}(19,7)
  \put(0,1.5){
  \begin{picture}(15,10)
  \fourpoints
  \qbezier[80](0,0)(5,7)(10,0)
  \put(5,0){\line(1,0){5}}
  \end{picture}}
\end{picture}
, \;
\begin{picture}(19,7)
  \put(0,1.5){
  \begin{picture}(15,10)
  \fourpoints
  \put(0,0){\line(1,0){5}}
  \put(5,0){\line(1,0){5}}
  \end{picture}}
\end{picture}
, \;
\begin{picture}(19,7)
  \put(0,1.5){
  \begin{picture}(15,10)
  \fourpoints
  \qbezier[40](5,0)(7.5,4)(10,0)
  \qbezier[40](5,0)(7.5,-4)(10,0)  
  \end{picture}}
\end{picture}
, \;
\begin{picture}(19,7)
  \put(0,1.5){
  \begin{picture}(15,10)
  \fourpoints
  \qbezier[40](0,0)(2.5,4)(5,0)
  \qbezier[40](0,0)(2.5,-4)(5,0)
  \end{picture}}
\end{picture}
, \;
\begin{picture}(19,7)
  \put(0,1.5){
  \begin{picture}(15,10)
  \fourpoints
  \qbezier[80](0,0)(5,7)(10,0)
  \qbezier[80](0,0)(5,-7)(10,0)
  \end{picture}}
\end{picture}
.
\end{center}
Note that there are only two different truly independent 
graphs in the sense that they are not a mere permutation of other graphs.
The first three graphs and the remaining three graphs form two
equivalence classes induced by permutations of $S_4$. 

Using the algorithm results in a maximum of $\tbinom{l+5}{l}$ terms that can appear. 
Combinatorial restrictions which we will discuss in the following can reduce this 
number, for instance $\eval{2211}$ for $l=3$ does not contain a $l_{34}^3$ term.

\subsection{Restrictions}\label{sec:restrictions}  
The analysis of the results we found shows that several restrictions reduce
the number of different terms that may appear in the end result. 

The first restriction naturally appears during the integration process.
In some cases our method for recursively constructing ``higher" correlators 
fails. It is not possible to repair this failure in a sensible manner by adding further
terms to the ansatz, but a simple identification of different functions $F$
immediately fixes the problem.

This behavior is a general property of the theory for $r\ge 3$, as we will see
in section \ref{sec:exact results}. For now it is sufficient to note that 
$F_{k_1-1,k_2,0,0} = F_{k_1,k_2-1,0,0}$, e.\,g., for $r=3$ we get
$F_{2100} = F_{1200}$  $+5$ more identifications by virtue of permutations.

The second restriction we encountered is the so called discrete symmetry of the correlators, which
limits the dimension of the kernel. By discrete symmetry we mean that a correlator 
which contains at least two fields of the same Jordan-level should be invariant 
under any transposition that exchanges these fields, for instance
\begin{align}
  P_{(12)} \eval{ \Psi_{k} \Psi_{k} \Psi_{k_3} \Psi_{k_4} } =  \eval{ \Psi_{k} \Psi_{k} \Psi_{k_3} \Psi_{k_4} } \; .
\end{align}
At this point we point out again that we have set, without loss of generality, 
all conformal weights $h_i$ to zero and wrote $\Psi_k$ instead of $\Psi_{(h,k)}(z)$.
In the next subsection we will discuss 
in more detail to what extent the above mentioned invariance limits the 
dimension of the kernel respectively show that in some cases no kernel term can show up at all.

The dimension of the kernel that finally shows up in the results is often smaller
than the one we would expect for the given logarithmic degree and given
discrete symmetry. The difference will show up especially if the logarithmic degree
is close to the maximum degree $l^{\mathrm{max}} = K-r+1$. 

The reason for this difference is that the ansatz does not allow all terms of $l_{ij}$
of a given degree to show up. For instance the correlator $\eval{2211}$
forbids the existence of terms of the type $l_{34}^3$ and by that limits the dimension
of the kernel of degree $3$. We also refer to this as combinatorial restriction, because
the restriction depends on the form of the correlator, e.\,g., the term $l_{34}^3$ is not
forbidden in $\eval{2221}$.

\subsection{Additional constants}\label{sec:additional constants}
As we have seen in section \ref{properties of O} the kernel of the operator $O$ is non-trivial.
That means that the results may come with additional constants. In order to understand the
meaning of these constants in the context of conformal field theory we rewrite the two basis terms
$K_1$, $K_2$ which every element of the kernel consists of as follows:
\begin{align}
  K_1 & := l_{12} + l_{34} - l_{13} - l_{24} = \log \left| \frac{z_{12} z_{34}}{z_{13} z_{24}} \right| \equiv \log|x| - \log |1-x| \; ,\\
  K_2 & := l_{12} + l_{34} - l_{14} - l_{23} = \log \left| \frac{z_{12} z_{34}}{z_{14} z_{23}} \right| \equiv \log |x| \; ,
\end{align}
where $x = \tfrac{z_{12}z_{34}}{z_{14}z_{32}}$ is the anharmonic ratio of the four points. 
The anharmonic ratio $x$ and its five possible involutions $\tfrac 1x$, $1-x$,
$1-\tfrac 1x$, $\tfrac{1}{1-x}$, and $\frac{x}{x-1}$
result in four linearly independent functions. If we take the 
logarithm of the absolute value of these four functions, then
we are left with only two independent solutions, namely $\log |x|$ and $\log|1-x|$.
The choice of the basis has no influence on the results and our choice of the basis
$K_1$, $K_2$ is given as above.

We can turn around the argument and ask for all functions of the anharmonic
ratio $x$, i.\,e., globally conformally invariant functions, which have the
additional property to be strictly polynomial in the $l_{ij}$. These 
functions are all in the kernel of the operator $O$. On the other hand,
there can be no other functions in the kernel if we restrict ourselves to
polynomials of the $l_{ij}$, since every member of the kernel must be
invariant under global conformal transformations and thus be a function
of $x$. This proves the statement in section \ref{properties of O}. 
However, we note that this yields only an upper bound on the size of the
kernel. We will see that the size may be reduced due to further symmetries.

Equation (\ref{eq:form of kernel}) gives us the maximal dimension of the kernel 
for logarithmic degree $l$,
\begin{align}
  K^{(l)} := & \left\{ \sum_{i=0}^l a_i K_1^i K_2^{d-i} : a_k \in \mathbb{R} \right\} \; , \label{eq:kernel space} \\
  d^{\mathrm{max}}(l) = & \;\; l+1 \; .
\end{align}
Up to a few exceptions we will notice that the full kernel never shows up
in any equation. These restrictions on the kernel are caused by 
the discrete symmetry and combinatorial constraints. Examples for 
combinatorial constraints are shown in the next two sections.

It is worth noting that a non-trivial kernel can show up in a correlator only if there 
is no primary field in the correlator present. This is obvious, since both kernel elements
$K_1$, $K_2$ refer to all four vertices $z_1, z_2, z_3,z_4$.

\subsubsection[Discrete symmetry for invariant $F$]{Discrete symmetry for invariant $\boldsymbol{F}$}\label{discrete symmetry for invariant F}
In this subsection we are interested in the impact on the kernel by a given symmetry. 
Since we consider four point correlation functions exclusively there are four interesting symmetry
groups only, namely $S_2$, $S_2 \times S_2$, $S_3$ and $S_4$.
Let us study an expression first, where the function $F$ is invariant under any
permutation, e.\,g., $(\ldots) F_{1111}$.

We start with the smallest symmetry group $S_2 = \{ 1, \perm{(12)} \}$, 
$P$ being, as usual, a permutation of the indices.
One immediately remarks that $\perm{(12)} K_1 = K_2$, $\perm{(12)} K_2 = K_1$ and
thus a $S_2$ invariant kernel of logarithmic degree $l$ has the form
\begin{align} \label{eq:s2 invariant kernel}
  K_{S_2}^{(l)} := \left\{ \sum_{i=0}^l a_i K_1^i K_2^{d-i} : a_k \in \mathbb{R}, a_k = a_{l-k} \right\}  \; .
\end{align}
Therefore the maximum number of constants $d^{\mathrm{max}}_{S_2}$ that could 
appear for logarithmic degree $l$ is 
\begin{align}
  d^{\mathrm{max}}_{S_2}(l) =  \left\lfloor \tfrac l2 \right\rfloor + 1 \; ,
\end{align}
where $\lfloor . \rfloor$ denotes the floor function.

If we replace the transposition $\perm{(12)}$ by $\perm{(34)}$ all statements
stay true. Thus when restricting to kernel space $K := \bigcup_{l} K^{(l)}$ we have
$\perm{(12)} \equiv \perm{(34)}$. This in turn means that the kernel is not only $S_2$ invariant, but 
automatically has full $S_2 \times S_2 = \{ 1, \perm{(12)}, \perm{(34)}, \perm{(12)(34)}\}$ invariance:
\begin{align}
  K_{S_2 \times S_2}^{(l)} \equiv & K_{S_2}^{(l)} \; , \label{eq:s2 equivs s2xs2} \\
  d^{\mathrm{max}}_{S_2 \times S_2}(l) = & \left\lfloor \tfrac l2 \right\rfloor + 1 \; . \label{eq:dim s2xs2 invariant kernel}
\end{align}
For the $S_3$ symmetry we note, that a $S_3$ invariance extends to $S_4$ invariance. 
This is because $S_3$ invariance in particular means $\perm{(12)}$ invariance 
which, as explained above, also means $\perm{(34)}$ invariance. By this we
immediately obtain full $S_4$ invariance:
\begin{align}
  K_{S_3}^{(l)} \equiv K_{S_4}^{(l)} \label{eq:s3 equivs s4} \; .
\end{align}
No linear combination of $K_1$, $K_2$ is $S_4$ invariant, but higher terms 
in $K_1$, $K_2$ have this property. The first two dimensional kernel $d=2$
can be found for logarithmic degree $l=6$:
\begin{align}\label{eq:s4 invariant kernel}
  K_{S_4}^{(2)} & := K_1^2 - K_1 K_2 + K_2^2 \; ,           \nonumber\\
  K_{S_4}^{(3)} & := (2 K_1 - K_2)(2 K_2-K_1) (K_1+K_2)  \; ,   \nonumber\\
  K_{S_4}^{(4)} & := (K_{S_4}^{(2)})^2 \; ,    \nonumber\\
  K_{S_4}^{(5)} & := K_{S_4}^{(3)} K_{S_4}^{(2)} \; ,     \nonumber\\
  K_{S_4}^{(6,(2,2,2))} & := (K_{S_4}^{(2)})^3 \; ,   \nonumber\\
  K_{S_4}^{(6,(3,3))} & :=  (K_{S_4}^{(3)})^2 \; , \nonumber\\
  K_{S_4}^{(7)} & := (K_{S_4}^{(2)})^2 K_{S_4}^{(3)} \; .
\end{align}
%
%
%
These results are unique, up to constants and linear combinations. Of course
any combination of the form $(K_{S_4}^{(2)})^i (K_{S_4}^{(3)})^j$ leads to a kernel of 
logarithmic degree $2i+3j$ and we believe that the kernel space is not 
larger than this, though it is not important since we consider kernels up to logarithmic 
degree $l=6$ only in the further course of this paper.

We expect the dimension of the $S_4$ invariant kernel to be the 
number of possible partitions of the of the degree in the numbers $2$ and $3$, e.\,g., 
$6=2+2+2$ as well $6=3+3$. This in turn means that every integer $6$ can be
represented in two different ways, leading to the following number of 
degrees of freedom that could appear at most for logarithmic degree $l$:
\begin{align}\label{eq:dim s4 invariant kernel}
  d^{\mathrm{max}}_{S_4}(l) = \left\{
  \begin{array}{ll}
    \left\lfloor\tfrac{l}{6}\right\rfloor  & \; ; l=6k+1, \; k \in \mathbb{N}_0 \\
   \left\lfloor\tfrac{l}{6}\right\rfloor+1 & \; ; \mathrm{else}\; .
  \end{array}\right. 
\end{align}
%

\subsubsection[Discrete symmetry and non-invariant $F$]{Discrete symmetry and non-invariant $\boldsymbol{F}$}
In the previous subsection we analyzed the structure of the kernel under
symmetry groups and found that we have to consider $S_2$ and $S_4$
symmetry groups only. This holds if the function $F_{\ldots}$ itself is invariant
under permutations. 

Things get more complicated if $F_a$ is mapped to $F_b$ by 
the permutation. For example we know that the $S_2$ invariant kernel
for one $F$ ($F_1$ for short) is one-dimensional, namely 
$K_{S_2}^{(1)} = \{ a(K_1+K_2) : a \in  \mathbb{R}\}$. But if we have two $F$,
which are related by the permutation, e.\,g., $F_{1022}$ and $F_{0122}$,
then the dimension of the kernel for $F_{1022}$ becomes larger. 
The kernel would be 
\[
  (c K_1 + c' K_2) F_{1022} + (c' K_1 + c K_2) F_{0122} \;, 
  \]
or $\mathcal{P}_{S_2} ( K^{(2)}  F_{1022})$ with \mbox{$\mathcal{P}_{S_2}=1 + \perm{(12)}$} for 
short. 

The kernel dimension therefore not only depends on the symmetry group
and the logarithmic degree, but also on the size of the equivalence
class of functions $F$ which are involved. The size of the equivalence class
is noted by $F_{n}$ and the results for the logarithmic degrees $l=1,2,\ldots 5$
are listed in appendix \ref{overview of kernel}.

The simple rule that $S_2$ corresponds to $S_2 \times S_2$ respectively
that $S_3$ corresponds to $S_4$ does not hold for $n>1$, therefore we have 
to discuss all four symmetry groups in the appendix. The dimension of the
kernel decreases with increasing size of the symmetry group and increases
with increasing size of the equivalence class. It is interesting though not 
surprising, that the full kernel $K^{(l)}$ is recovered, if the size of the equivalence
class $|F|$ equals the size of the symmetry group $|S|$.

\section{Results for Jordan-rank $r=2$}\label{sec:results for r=2}
In this section we present and discuss the results for a logarithmic conformal field theory 
with Jordan-level $r=2$. We have used the algorithm described in subsection \ref{sec:algorithm}
to obtain these results and though known, e.\,g. \cite{Flo00}, we can write them in a more
appealing form. Also we will discuss the appearance of an additional degree of freedom,
which shows up for $\eval{1111}$.

We start with simply writing down the first three expressions that our algorithm provides:
\begin{align}
  \eval{1000} = & F_{0} \; , \\ 
  \eval{1100} = & \mathcal{P}_{S_2} \Big\{ \tfrac{1}{2}  F_{1100} - \diagAone F_{0} \Big\} \; , \nonumber\\
  \eval{1110} = & \mathcal{P}_{S_3} \Big\{ \tfrac{1}{6}  F_{1110}+(\tfrac{1}{2} \perm{(13)}\!-1)\diagAone F_{0110}+ \big[ \diagBtwo-\tfrac{1}{2} \diagCtwo \big] F_{0} \Big\} \; . \nonumber\\
\end{align}
with $\mathcal{P}_{X} = \sum_X \perm{(x)}$. Writing the results this way makes the
discrete symmetry manifest, that is $S_2$ for $\eval{1100}$ and $S_3$ invariance for $\eval{1110}$.

For the correlator $\eval{1111}$ we have a logarithmic partner field at every vertex
which means that we can expect getting a non-trivial kernel for the first time. The result
without kernel is given by
\begin{align}
  \eval{1111} = & \mathcal{P}_{S_4} \Big\{  \tfrac{1}{24}  F_{1111} + (\tfrac{1}{6} \perm{(13)}-\tfrac{1}{3})\diagAone F_{0111} + \nonumber\\
  & \phantom{\mathcal{P}_{S_4} \Big\{} \big[ \tfrac{1}{2}  (\perm{(24)} - 1)\diagAtwo+( 1 -\tfrac{1}{2} \perm{(14)})\diagBtwo - \tfrac{1}{4}  \diagCtwo \big] F_{0011} + \nonumber\\
  & \phantom{\mathcal{P}_{S_4} \Big\{} \big[ \tfrac{1}{2} \diagEthree+\tfrac{1}{3}  \diagFthree- \diagCthree \big] F_{0} \big\}  \; .
\end{align}
the contribution to the kernel is
\begin{align}
  \mathrm{Ker}_{\eval{1111}} = \mathcal{P}_{S_4} \Big\{ K_{S_2}^{(2)} F_{0011} \Big\} \; .
\end{align}
That we get a two-dimensional kernel for $F_{0011}$ is not surprising, since there are $6$ functions
$F$ belonging to the equivalence class of $F_{0011}$ and the resulting $K_{S_2}^{(2)}$ can be
read of the table for logarithmic degree $2$ from the appendix. 

The inverse question is more interesting, namely we are interested in understanding, why
no other kernel term shows up at all. For logarithmic degree $l=1$ the equivalence class of 
$F_{0111}$ is four and thus there is no kernel term showing up. According to 
(\ref{eq:dim s4 invariant kernel}) the $S_4$ invariant kernel of logarithmic degree $l=3$
should be one-dimensional. We can immediately understand why this kernel term does
not show up, by looking at the graphical representation:
\begin{align}\label{eq:graph repres K_S4^3}
  K_{S_4}^{(3)} = & \mathcal{P}_{S_4} \Big\{  \tfrac{1}{2}\diagDthree + 2 \diagBthree - 3 \diagAthree - 3 \diagCthree + \nonumber\\
                    & \qquad\; \tfrac{3}{2} \diagEthree+2 \diagFthree \Big\} \; .
\end{align}
This shows us that terms of the form $l_{12}^3$ appear, which is impossible for a 
Jordan-rank $r=2$ theory. Though three free legs are available the
three-fold connection between vertices $i$ and $j$ is forbidden for $r=2$.
Of course higher Jordan-rank LCFT $r>2$ are allowed to include such terms,
but similar combinatorial restrictions will show up for $r=3$ as well.

\section{Results for Jordan-rank $r=3$}
While the general structure of the correlators for Jordan-rank $r=2$ has been known before,
nobody so far has studied the form of correlators for LCFTs beyond the case of $r=2$.
With what we have learned we can apply our methods to the case $r=3$ in order to determine
the form of all correlators for a theory of Jordan-rank $r=3$.

Analogously to $r=2$ the starting point for the recursion is given by
\begin{align}
  \eval{1100} = F_0 \quad , \quad  \eval{2000} = & F_0 \; .
\end{align}
The missing correlators result from applying a permutation 
to the correlators.
\begin{align}
  \eval{0012} = & \mathcal{P}_{S_2} \Big\{ \tfrac{1}{2} F_{0012} - \diagAOneMod F_0 \Big\} \; , \\
  \eval{1110} = & \mathcal{P}_{S_3} \Big\{ \tfrac{1}{6} F_{1110} - \tfrac 12 \diagAone F_{0} \Big\} \; , \\
  \eval{2200} = & \mathcal{P}_{S_2 \times S_2} \Big\{ \tfrac{1}{4}  F_{2200} - \tfrac{1}{2}  \diagAone F_{1200}+\tfrac{1}{2}  \diagCtwo F_{0} \Big\} \; ,\\
  \eval{1120} = & \mathcal{P}_{S_2} \Big\{ \tfrac{1}{2}  F_{1120}   - (1 \!+\! \perm{(23)}\!+\!\perm{(13)}) \diagAone F_{0120}+(\tfrac{1}{2} \!-\!\perm{(23)})\diagAone F_{1110} + \nonumber\\
    &  \big[ (1+\tfrac{3}{2} \perm{(23)} )\diagBtwo - (\tfrac{1}{4}  + \tfrac{1}{2} \perm{(23)})\diagCtwo \big] F_{0} \Big\} \label{eq:corr1120-r=3}
\end{align}
where as before $\mathcal{P}_{X} = \sum_{x \in X} \perm{x}$.

The above correlators do not have an additional degree of freedom because
they contain at least one primary field. The simplest correlator with no primaries is
\begin{align}
  \eval{1111} = \mathcal{P}_{S_4} \Big\{
    \tfrac{1}{24}  F_{1111} + (\tfrac{1}{6} \perm{(13)} -\tfrac{1}{3})\diagAone F_{0111} + (\tfrac{1}{4} \diagAtwo) F_{0} \Big\} \; .
\end{align}
This is the first correlator for $r=3$ which has a non-trivial kernel, namely
\begin{align}
  \mathrm{Ker}_{\eval{1111}} = c_1 K_{S_4}^{(2)} F_{0} \; .
\end{align}
The restriction that the expression needs to be invariant under $S_4$ permutations is very strong
and forbids any kernel terms of degree one to show up.  

The remaining correlators containing at least a primary field are
\begin{align}
  \eval{2210} = & \mathcal{P}_{S_2} \Big\{  \tfrac{1}{2}  F_{2210}+(\tfrac{1}{2} \!-\!\perm{(13)})\diagAone F_{2200} - (1 \!+\! \perm{(23)}\!-\!\perm{(13)})\diagAone F_{1210}+ \nonumber\\
  & \big[ 2 \diagBtwo- \diagCtwo \big] F_{1200}+ \big[ \perm{(23)}\diagBtwo+(\tfrac{1}{2} \!-\!\perm{(13)}) \diagCtwo\big] F_{1110}+ \nonumber\\
  & \big[ (- 1 -\perm{(23)}+\perm{(12)})\diagBtwo+ \tfrac{1}{2} (1 + \perm{(23)}+ \perm{(13)})\diagCtwo \big] F_{0120}+ \nonumber\\
  & \big[ (\perm{(13)}-2)\diagAthree+\tfrac{1}{2}  \diagDthree- \diagFthree \big] F_{0} \Big\}
\end{align}
and
\begin{align}
  \eval{2220} = & \mathcal{P}_{S_3} \Big\{ \tfrac{1}{6}  F_{2220}+ (\tfrac{1}{2} \perm{(13)} - 1)\diagAone F_{1220}+ \nonumber\\
  & \big[ \perm{(23)}\diagBtwo+(\tfrac{1}{2} -\perm{(23)})\diagCtwo \big] F_{1120}+ \nonumber\\
  & \big[ (\tfrac{1}{2} \perm{(12)}-1)\diagBtwo+(\tfrac{1}{2} +\tfrac{1}{4} \perm{(13)})\diagCtwo \big] F_{0220}+\nonumber\\
  & \big[ \tfrac{1}{2}  \diagDthree - \diagAthree +\tfrac{1}{3} \diagFthree \big] F_{1110}+ \nonumber\\
  & \big[ (2\perm{(13)}-1)\diagAthree-\tfrac{1}{2} \perm{(13)}\diagDthree-  \diagFthree\big] F_{0120}+ \nonumber\\
  & \big[ \tfrac{1}{2}  \diagCfour+\tfrac{1}{8}  \diagEfour + \tfrac{3}{4}  \diagFfour-  \diagGfour \big] F_{0} \Big\} \; .
\end{align}

Finally there are, up to permutations, four correlators without primary field and at least one field
being of Jordan-level 2.
\begin{align}
  \eval{1112} = & \mathcal{P}_{S_3} \Big\{  \tfrac{1}{6}  F_{1112} +
       (\tfrac{1}{6} -\tfrac{1}{3} \perm{(14)})\diagAone F_{1111} + \nonumber\\
  & (\perm{(13)(24)}-\perm{(24)}-\tfrac{1}{2} \perm{(13)})\diagAone F_{0112} + \nonumber\\
  & \big[ (\tfrac{1}{12} -\tfrac{1}{6} \perm{(34)}+\tfrac{1}{4} \perm{(24)})\diagBtwo-\tfrac{1}{6} \diagAtwo+\tfrac{1}{12} \perm{(14)}\diagCtwo \big] F_{1110} + \nonumber\\
  & \big[ ( 1 +\perm{(34)}-\perm{(14)})\diagBtwo+ (\perm{(24)}- 1 )\diagAtwo - \tfrac{1}{2} \diagCtwo \big] F_{0012} + \nonumber\\
  & \big[ (\tfrac{2}{3}  -\tfrac{1}{3} \perm{(24)})\diagAtwo+( \tfrac{2}{3} \perm{(34)}-\tfrac{2}{3}+\tfrac{2}{3} \perm{(24)}-\tfrac{1}{3} \perm{(124)})\diagBtwo+ \nonumber\\
  & \phantom{\big[} (\tfrac{1}{6} \perm{(13)}-\tfrac{1}{3} \perm{(13)(24)})\diagCtwo \big] F_{0111} + \nonumber\\
  & \big[ ( \tfrac{1}{4}  -\tfrac{1}{4} \perm{(34)}+\tfrac{3}{4} \perm{(24)}+\tfrac{3}{4} \perm{(14)})\diagAthree  - \tfrac{1}{4} \perm{(24)}\diagDthree    -\tfrac{1}{6}  \diagFthree + \nonumber\\
  & \phantom{\big[} (\tfrac{1}{2} -\tfrac{3}{2} \perm{(24)})\diagCthree - \tfrac{1}{2} (1  + \perm{(24)})\diagBthree + \tfrac{1}{4}(1  - \perm{(14)})\diagEthree \big] \Big\} F_{0}
\end{align}
This correlator has 6 additional degrees of freedom:
\begin{align}
  \mathrm{Ker}_{\eval{1112}} = \mathcal{P}_{S_3} \Big\{ & c_1 (2K_2 - K_1) F_{0112}  +
    K_{S_4}^{(2)} F_{1110} + \nonumber\\
  & \big[ c_3 K_1^2 + c_4 (K_2^2 - K_1 K_2) \big] F_{0111}  + \nonumber\\
  & \big[ c_5 K_1^2 + c_6 (K_2^2 - K_1 K_2) \big] F_{1002} \Big\}  \; .
\end{align}
The set $\{ F_{0112}, F_{1012}, F_{1102}\}$ allows a one dimensional kernel, namely
\mbox{$2 K_2 - K_1$}. Note that this kernel is not $S_3$ invariant. 

For the self-invariant terms $F_{1110}$ and $F_0$
we remarked in subsection \ref{discrete symmetry for invariant F} that 
$S_3$ invariance implies $S_4$ variance. The dimension of the 
$S_4$-invariant kernel is $n_{\mathrm{max}}^{S_4} = 1$ for $d=2, 3$. The corresponding 
kernel term for $F_{1110}$ shows up, combinatorial restrictions forbid the same for $F_0$. 
The only kernel of degree $3$ that would have been possible is $K_{S_4}^{(3)}$, but this one 
includes $l_{ij}^3$ terms for all $1 \le i < j \le 4$, which is not compatible with the contraction 
rules as described in subsection \ref{sec:algorithm}--$\eval{2111}$ cannot 
contain $l_{34}^3$ terms, cf. \ref{eq:graph repres K_S4^3}.

For the 3-element sets $\{ F_{0111}, F_{1011}, F_{1101} \}$ respectively 
$\{ F_{0012}, F_{0102}, F_{1002} \}$ we know from the kernel analysis in 
appendix \ref{overview of kernel} that there is a two-dimensional kernel.

It should be noted that $\eval{1211}$ is generated by applying $\perm{(12)}$ to $\eval{2111}$. 
The same holds for the additional terms of the kernel. This means that the degrees of freedom
we have for $\eval{2111}$ are not available for the permutations of this correlator, e.\,g., $\eval{1211}$.

The correlator $\eval{2211}$ comes with a high number of additional degrees of freedom,
some of these are restricted by combinatorial constraints. The correlator without kernel
terms has the form
\begin{align}
  \eval{2211} = & \mathcal{P}_{S_2 \times S_2} \Big\{ \tfrac{1}{4}  F_{2211} + 
    (\tfrac{1}{2}-\perm{(13)})\diagAone F_{2201} + \nonumber\\
  & ( \tfrac{1}{2} \perm{(13)(24)} -\perm{(23)} ) \diagAone F_{1211} +\nonumber\\
  & \big[ \tfrac{1}{2} \perm{(23)}\diagAtwo + (\tfrac{1}{2} \perm{(14)}-1)\diagBtwo+\tfrac{1}{4} \diagCtwo \big] F_{2200} + \nonumber\\
  & \big[ (\tfrac{1}{2} \perm{(23)}-\tfrac{1}{6})\diagAtwo+(\tfrac{1}{3} -\tfrac{1}{2} \perm{(14)})\diagBtwo+\tfrac{1}{12} \diagCtwo \big] F_{1111} + \nonumber\\
  & \big[ (\perm{(243)}+\perm{(24)}+\perm{(12)}-\perm{(12)(34)}-\perm{(124)}+\perm{(142)})\diagBtwo + \nonumber\\
  & \phantom{\big[} ( 1 -\perm{(23)})\diagAtwo - \perm{(24)}\diagCtwo \big] F_{1201}+ \nonumber\displaybreak[0]\\
  & \big[ (\tfrac{1}{2} +\tfrac{1}{2} \perm{(23)}-\tfrac{1}{2} \perm{(24)}+\tfrac{1}{2} \perm{(12)}+\tfrac{1}{2} \perm{(124)}+\tfrac{3}{4} \perm{(142)}+\tfrac{1}{4} \perm{(14)})\diagBtwo + \nonumber\\
  & \phantom{\big[} (-\tfrac{1}{2} -\perm{(24)})\diagAtwo - (\tfrac{1}{4} + \tfrac{1}{2} \perm{(14)})\diagCtwo \big] F_{0211} + \nonumber\displaybreak[0]\\
  & \big[ 2\perm{(12)}\diagAthree - \perm{(12)} \diagBthree - \perm{(13)}\diagCthree - \tfrac{1}{2} \diagDthree \big] F_{1200}+\nonumber\\
  & \big[ \tfrac{1}{2} (\perm{(34)} - 1 +\perm{(243)}- \perm{(24)} -2\perm{(1243)} - \perm{(124)} -2 \perm{(143)} - \perm{(14)})\diagAthree +  \nonumber\\
  & \phantom{\big[} ( 1 +\perm{(24)})\diagBthree -  (\tfrac{1}{2} + \tfrac{1}{2} \perm{(24)})\diagEthree+\perm{(34)}\diagFthree + \nonumber\\
  & \phantom{\big[} (\perm{(34)}+\perm{(243)}-\perm{(12)(34)} - \perm{(13)} )\diagCthree + \tfrac{1}{2} \perm{(14)}\diagDthree \big] F_{0102} + \nonumber\displaybreak[0]\\
  & \big[ \tfrac{1}{3} (\perm{(23)} - 2 + \perm{(34)}+ 2 \perm{(234)} - \perm{(134)}+ 4 \perm{(13)}+ 3 \perm{(14)} - \perm{(143)})\diagAthree+ \nonumber\\
  & \phantom{\big[} (\tfrac{1}{6} -\tfrac{1}{3} \perm{(13)})\diagDthree -(\tfrac{4}{3} +\tfrac{2}{3} \perm{(23)})\diagBthree+ ( \tfrac{1}{3} \perm{(34)}-\tfrac{4}{3})\diagFthree + \nonumber\\
  & \phantom{\big[} (\tfrac{2}{3} \perm{(23)}-2\perm{(243)}+\tfrac{4}{3} \perm{(13)})\diagCthree - (\tfrac{1}{3} + \tfrac{1}{3} \perm{(23)}+\perm{(24)})\diagEthree \big] F_{1101} + \nonumber\displaybreak[0]\\
  & \big[ (\tfrac{2}{3} \perm{(23)}-\tfrac{5}{4}+\tfrac{7}{12} \perm{(24)}-\tfrac{7}{12} \perm{(12)}+\tfrac{3}{4} \perm{(124)}+\tfrac{7}{12} \perm{(132)}+\tfrac{2}{3} \perm{(142)}+\tfrac{5}{4} \perm{(13)} + \nonumber\\
  & \phantom{\big[} \tfrac{1}{12} \perm{(1423)}-\tfrac{1}{6} \perm{(14)}-\tfrac{1}{12} \perm{(14)(23)})\diagAthree -(\tfrac{4}{3} \perm{(24)}+\tfrac{2}{3} \perm{(12)})\diagBthree + \nonumber\\
  & \phantom{\big[} (\tfrac{5}{12} -\tfrac{5}{12} \perm{(24)}-\tfrac{1}{4} \perm{(14)})\diagDthree - (\tfrac{4}{3}  +\tfrac{2}{3} \perm{(24)}+\tfrac{1}{2} \perm{(14)})\diagFthree  + \nonumber\\
  & \phantom{\big[} (\tfrac{4}{3} \perm{(23)}-\tfrac{7}{6}+\tfrac{7}{6} \perm{(24)}+\tfrac{1}{6} \perm{(12)}+\tfrac{7}{6} \perm{(132)})\diagCthree+ \nonumber\\
  & \phantom{\big[} (\tfrac{11}{12} -\tfrac{5}{4} \perm{(24)}-\tfrac{7}{12} \perm{(13)}+\tfrac{1}{4} \perm{(13)(24)})\diagEthree \big] F_{0111} + \nonumber\displaybreak[0]\\
  & \big[  (\perm{(24)}-\tfrac{1}{2} \perm{(23)}-\tfrac{3}{4} \perm{(14)})\diagCfour + \tfrac{1}{2} (\tfrac{1}{2} - \perm{(23)}+ \perm{(24)} -\tfrac{1}{4} \perm{(14)})\diagFfour+\nonumber\\
  & \phantom{\big[} \tfrac{1}{2} ( 1 +  \perm{(23)} - 4\perm{(24)} - \perm{(14)} - 4 \perm{(13)})\diagDfour - (\tfrac{1}{8} - \tfrac{5}{8} \perm{(24)})\diagIfour + \nonumber\\
  & \phantom{\big[} \tfrac{1}{2} ( 1+ \perm{(23)}  - \perm{(14)})\diagAfour+(\tfrac{3}{4} \perm{(24)}-\perm{(23)}+\tfrac{5}{4} \perm{(13)})\diagBfour+\nonumber\\
  & \phantom{\big[} (\tfrac{1}{2} \perm{(23)}-\tfrac{1}{2} \perm{(24)}- 1 )\diagGfour+\tfrac{1}{2} \perm{(23)}\diagHfour+ \tfrac{3}{16}\diagEfour + \nonumber\\
  & \phantom{\big[}  (2\perm{(13)}+\perm{(14)}-\perm{(13)(24)})\diagJfour+ \diagKfour \big] F_{0} \Big\} \; ,
\end{align}
where $\mathcal{P}_{S_2 \times S_2} = 1 + \perm{(12)} + \perm{(34)} + \perm{(12)(34)}$.
The kernel of $\eval{2211}$ has a dimension of 18:
\begin{align}
  \mathrm{Ker}_{\eval{2211}} = & \mathcal{P}_{S_2 \times S_2} \Big\{   
    K_{S_2}^{(1, d=1)} F_{2201} + K_{S_2}^{(1, d=1)} F_{1211}  + \nonumber\\
  & K^{(2, d=3)} + K_{S_2}^{(2, d=2)} (F_{2200} + F_{0211} + F_{1111}) +  \nonumber\\
  & (K_1- K_2)^2 (K_1 + K_2)  ( F_{1101} + F_{0111} + F_{1200}) + \nonumber\\
  & \big[ c (K_1-K_2)^2 K_1 + c' (K_1-K_2)^2 K_2  + c''(K_1-K_2) K_1 K_2 \big]  F_{0102} +\nonumber\\
  & K_{S_4}^{(4,d=1)} F_0 \Big\} \; ,
\end{align}
where we used a somewhat condensed notation and left out
almost all constants. If multiple $F$ in brackets show up you should add
the necessary constants
in your mind, for instance, $K_{S_2}^{(2, d=2)} (F_{2200} + F_{0211} + F_{1111})$
stands for $2 \cdot 3 = 6$ degrees of freedom. For better orientation
we added a small "$d=\ldots$" index to the common kernel terms which notes their
dimension.

For the sets $\{ F_{2201}, F_{2210}\}$ and $\{ F_{1211}, F_{2111} \}$ we get the expected 
one-dimen\-sional kernel $K_{S_2}^{(1)}$. Also the results for logarithmic degree 2 are not
surprising.

Things are more complicated for higher logarithmic degrees. For $d=3$ we would
have been expecting a two-dimensional kernel $K_{S_2}^{(3)}$ for $F_{1101}$,
$F_{1200}$, $F_{0111}$ and a full 4-dimensional kernel $K^{(3)}$ for $F_{0102}$.
But here we have to take into account the combinatorial restrictions again.
The basis elements of $K_{S_2}^{(3)}$, $K_{S_2}^{(3,a)} \sim K_1^3 + K_2^3$ 
contains $l_{12}^3, l_{13}^3, \ldots, l_{34}^3$ contributions and
$K_{S_2}^{(3,b)} \sim K_1^2 K_2 + K_1 K_2^2$ comes with $l_{12}^3, l_{34}^3$ terms.
Thus both basis elements are not allowed due to the $l_{34}^3$ term, but a linear
combination is, namely $(K_1-K_2)^2 (K_1+K_2)$, which contains no $l_{34}^2$ term.
For the set $\{ F_{0102} \equiv F_{0201}, F_{1002} \equiv F_{2001}, 
F_{0120} \equiv F_{0210}, F_{1020}\equiv F_{2010}$ the four dimensional kernel
reduces by the combinatorial constraint to a three dimensional one.

The reasoning for $d=4$ goes along the same line. We expect from (\ref{eq:dim s2xs2 invariant kernel})
a three dimensional kernel space, but also we have two restrictions. No $l_{ij}^4$ term may show up,
not even $l_{12}^4$, because of the $S_2 \times S_2$ invariance. And no $l_{34}^3$ term is allowed.
These two restrictions limit the kernel to $K = K_1 K_2 (K_1-K_2)^2$, leaving us with a one-dimensional kernel.
\begin{align}
  \eval{2221} = &  \mathcal{P}_{S_3} \Big\{ \tfrac{1}{6}  F_{2221} + (\tfrac{1}{6} -\tfrac{1}{3} \perm{(14)})\diagAone F_{2220} + ( \tfrac{1}{2} \perm{(13)}- 1)\diagAone F_{1221} + \nonumber\\
  & \big[ (2 -\perm{(34)}-\tfrac{1}{2} \perm{(12)}+2\perm{(12)(34)}+\perm{(14)})\diagBtwo - (1 + \perm{(24)})\diagAtwo + \nonumber\\
  & \phantom{\big[} (-\tfrac{1}{2} -\tfrac{3}{4} \perm{(13)})\diagCtwo \big] F_{0221} + \nonumber\\
  & \big[ (2 -\perm{(24)})\diagAtwo+(2 -2\perm{(34)}-2\perm{(12)}+2\perm{(12)(34)}-\perm{(124)})\diagBtwo+\nonumber\\
  & \phantom{\big[} (\perm{(13)(24)}-\tfrac{1}{2} \perm{(13)})\diagCtwo \big] F_{1220}+\nonumber\\
  & \big[ (1-\tfrac{1}{2} \perm{(23)}+\tfrac{1}{2} \perm{(243)} +\perm{(34)}+\perm{(234)}+\perm{(24)}-\perm{(14)})\diagBtwo - \diagAtwo +\nonumber\\
  & \phantom{\big[} (-\perm{(24)}-\tfrac{1}{2} \perm{(13)(24)})\diagCtwo \big] F_{1121}+ \nonumber\displaybreak[0]\\
  & \big[ \tfrac{2}{3} \diagCthree -\tfrac{2}{3} \diagBthree-\tfrac{2}{3} \diagEthree-\tfrac{1}{3} \diagFthree - \tfrac{1}{6} \diagDthree + \nonumber\\
  & \phantom{\big[} (\tfrac{1}{3} +\tfrac{2}{3} \perm{(34)})\diagAthree  \big] F_{1111}+ \nonumber\displaybreak[0]\\
  & \big[\tfrac{1}{2}  \diagDthree-2\perm{(23)}\diagCthree+\perm{(23)}\diagEthree+ \diagFthree + \nonumber\\
  & \phantom{\big[} (\perm{(134)}-\perm{(34)}-\perm{(13)})\diagAthree \big] F_{1120} + \nonumber\displaybreak[0]\\
  & \big[ \tfrac{1}{2}(1 - \perm{(34)}- 2 \perm{(13)}-\perm{(134)})\diagAthree+ \diagBthree+( 1 -\perm{(12)})\diagCthree+ \nonumber\\
  & \phantom{\big[} \tfrac{1}{4} \perm{(13)}\diagDthree-\tfrac{1}{2} \diagEthree+\tfrac{1}{2} \diagFthree \big] F_{0220} + \nonumber\displaybreak[0]\\
  & \big[ \tfrac{1}{2} (\perm{(23)}-2 \perm{(34)}+2 \perm{(234)} - \perm{(243)}+\perm{(13)(24)} +\perm{(12)(34)}+3 \perm{(1243)}+ \perm{(14)(23)}+  \nonumber\\
  & \phantom{\big[}  3 \perm{(124)}  -\!\perm{(12)} +\! 2 \perm{(132)} + \!\perm{(1342)} +\!\perm{(24)} -\!\perm{(1324)}+ \!4 \perm{(143)}-\!\perm{(1423)})\diagAthree+ \nonumber\\
  & \phantom{\big[} ( 1 -2\perm{(34)}+\perm{(23)}+2\perm{(234)}+\perm{(243)}-\perm{(12)}+2\perm{(12)(34)}+\perm{(1243)}-\perm{(13)}+ \nonumber\\
  & \phantom{\big[}  2\perm{(132)})\diagCthree  - (1 +2\perm{(34)} + \perm{(24)} + \perm{(14)})\diagFthree + \nonumber\\
  & \phantom{\big[}(\tfrac{1}{2} -\tfrac{3}{2} \perm{(23)}-\tfrac{1}{2} \perm{(24)}-\tfrac{3}{2} \perm{(14)})\diagEthree  - (\tfrac{1}{2} \perm{(23)}+\perm{(14)})\diagDthree +  \nonumber\\
  & \phantom{\big[} (-\perm{(23)}-2\perm{(24)}-\perm{(12)})\diagBthree \big] F_{0121} + \nonumber\displaybreak[0]\\
  & \big[ \tfrac{2}{3}  \diagBfour -\tfrac{2}{3} \perm{(14)}\diagAfour - (\tfrac{1}{3} +\perm{(24)})\diagCfour+ (\tfrac{2}{3} \perm{(34)}-\!\tfrac{1}{3})\diagFfour+ \nonumber\\
  & \phantom{\big[} (\tfrac{4}{3} +\tfrac{2}{3} \perm{(34)})\diagDfour-\tfrac{1}{6}  \diagEfour+\tfrac{2}{3}   \diagGfour-  \diagHfour+\nonumber\\
  & \phantom{\big[} (\tfrac{4}{3} \perm{(14)}-\tfrac{2}{3} )\diagJfour -\tfrac{1}{3} \perm{(24)}\diagIfour - \tfrac{2}{3}  \diagKfour \big] F_{1110} +\nonumber\displaybreak[0]\\
  & \big[  (\tfrac{1}{4} \perm{(34)}+\tfrac{1}{2} \perm{(23)}-\tfrac{1}{2} \perm{(234)}-\tfrac{1}{4} +\tfrac{1}{4} \perm{(14)})\diagFfour+ (\perm{(23)}-\perm{(34)})\diagKfour + \nonumber\\
  & \phantom{\big[} (\perm{(23)}-\tfrac{1}{2}+\tfrac{1}{2} \perm{(243)}+\tfrac{1}{2} \perm{(24)})\diagCfour + (\perm{(23)}-1)\diagBfour+ \diagHfour+ \nonumber\\
  & \phantom{\big[} \tfrac{1}{2} ( \perm{(234)} - 3 \perm{(34)} - 3 - \perm{(23)} + 2 \perm{(13)} - 2 \perm{(134)}+ \perm{(143)} - \perm{(14)})\diagDfour+ \nonumber\\
  & \phantom{\big[} (\tfrac{1}{2} \perm{(234)}- 1 -\perm{(34)}-\tfrac{1}{2} \perm{(23)}+\tfrac{1}{2} \perm{(13)}-\tfrac{1}{2} \perm{(134)})\diagGfour+ \tfrac{3}{8} \diagEfour+ \nonumber\\
  & \phantom{\big[} (\tfrac{3}{2} \perm{(23)}-\tfrac{1}{2}-\tfrac{1}{2} \perm{(143)}+\tfrac{1}{2} \perm{(14)})\diagAfour+  (\tfrac{3}{4}   +\tfrac{1}{4} \perm{(24)})\diagIfour+ \nonumber\\
  & \phantom{\big[} ( 1 -\perm{(13)}-\perm{(14)}+\perm{(1324)})\diagJfour  \big] F_{0012} + \nonumber\displaybreak[0]\\
  & \big[ \tfrac{1}{2} ( \perm{(34)}+\perm{(23)}- 3 \perm{(243)}- \perm{(12)(34)}-6\perm{(123)}-6\perm{(1234)}+\perm{(132)}-2 \perm{(13)(24)}+ \nonumber\\
  & \phantom{\big[} 2 \perm{(1324)}-3 \perm{(1432)}-2 \perm{(14)}+2 \perm{(1423)})\diagDfour+(\tfrac{1}{2} \perm{(24)}+\tfrac{3}{2} \perm{(13)})\diagHfour+ \nonumber\\
  & \phantom{\big[} \tfrac{1}{2} (3 \perm{(24)} -\! \perm{(23)}-\!3 \perm{(234)}-\! \perm{(12)}\!+\! \perm{(12)(34)}\!+\! \perm{(123)})\diagBfour+\tfrac{3}{2} \perm{(24)}\diagIfour+ \nonumber\\
  & \phantom{\big[} \tfrac{1}{2} (2 \perm{(23)} -\! \perm{(34)} +\! \perm{(12)(34)} - \!\perm{(132)} -\! \perm{(14)})\diagAfour+ \tfrac{1}{2} (\perm{(243)}-\! \perm{(23)}-\! \perm{(234)}+ \nonumber\\
  & \phantom{\big[} \perm{(124)})\diagCfour+ (2 \perm{(1234)}- \perm{(23)}- \perm{(234)}+\perm{(243)}- \perm{(124)})\diagFfour+ \nonumber\\
  & \phantom{\big[} \tfrac{1}{2}  (\perm{(124)}+\perm{(13)}-\perm{(134)}- \perm{(13)(24)}-\perm{(14)}+ \perm{(1423)})\diagGfour+\nonumber\\
  & \phantom{\big[} ( 1 +\perm{(34)}+\perm{(24)}+\perm{(13)})\diagJfour+(2 -\perm{(34)})\diagKfour \big] F_{0120} + \nonumber\displaybreak[0]\\
  & \big[ \tfrac{1}{3} (\tfrac{17}{2} \perm{(34)}\!-\!1\!+\!\perm{(12)}\!+\!\tfrac{11}{2} \perm{(124)}\!+\!\tfrac{5}{2} \perm{(1342)}\!+\!2 \perm{(13)}\!-\! \perm{(13)(24)}\!+\!\tfrac{3}{2} \perm{(14)})\diagDfour + \nonumber\\
  & \phantom{\big[} (\tfrac{1}{2}   -\tfrac{5}{3} \perm{(34)}+\tfrac{2}{3} \perm{(24)}+\tfrac{1}{2} \perm{(12)(34)}+\tfrac{1}{3} \perm{(13)}-\tfrac{1}{3} \perm{(12)})\diagBfour+\tfrac{1}{3} \diagKfour+\nonumber\\
  & \phantom{\big[} (\tfrac{5}{6}   -\tfrac{1}{2} \perm{(34)}-\tfrac{5}{6} \perm{(24)}-\tfrac{2}{3} \perm{(12)}+\tfrac{2}{3} \perm{(12)(34)}-\tfrac{1}{6} \perm{(124)}+\tfrac{1}{3} \perm{(14)})\diagCfour+\nonumber\\
  & \phantom{\big[} (\tfrac{5}{6} \perm{(34)}-\tfrac{5}{6} +\tfrac{1}{3} \perm{(134)}+\tfrac{1}{2}\perm{(124)}-\tfrac{1}{2} \perm{(13)(24)}+\tfrac{1}{3} \perm{(143)}-\tfrac{1}{3} \perm{(14)})\diagGfour+ \nonumber\\
  & \phantom{\big[} (\tfrac{1}{2} \perm{(12)(34)}-\tfrac{1}{2} \perm{(34)}-\tfrac{7}{6} \perm{(12)}+\tfrac{5}{6} \perm{(142)})\diagAfour -\tfrac{1}{12} \perm{(13)}\diagEfour + \nonumber\\
  & \phantom{\big[} \tfrac{1}{3} (1 -\perm{(34)}- \perm{(24)}- \perm{(13)}-2 \perm{(134)}-4 \perm{(14)})\diagJfour -\tfrac{11}{6} \diagIfour+ \nonumber\\
  & \phantom{\big[} ( 1 -\perm{(34)}+\tfrac{1}{2} \perm{(24)}+\tfrac{1}{6} \perm{(134)}-\tfrac{1}{2} \perm{(124)})\diagFfour +\nonumber\\
  & \phantom{\big[} (-\tfrac{5}{6}  -\tfrac{1}{3} \perm{(13)}-\tfrac{1}{2} \perm{(13)(24)})\diagHfour \big] F_{0111} + \nonumber\displaybreak[0]\\
  & \big[ (\tfrac{1}{4} \!-\!\perm{(24)}+\tfrac{3}{4} \perm{(34)})\diagAfive + (1\!-\!\perm{(34)} )\diagBfive+(\perm{(24)}\!-\!\perm{(34)})\diagCfive+ \nonumber\\
  & \phantom{\big[} (\tfrac{3}{4} \perm{(24)}\!+\tfrac{5}{4}  )\diagIfive-(\tfrac{3}{4} +\tfrac{1}{4} \perm{(24)} )\diagHfive+(1\!-\!\perm{(34)}\!-\!2\perm{(14)})\diagJfive+ \nonumber\\
  & \phantom{\big[} (\perm{(24)}-\!\tfrac{7}{4}  +\tfrac{3}{4} \perm{(34)}-\!\perm{(14)})\diagGfive +(\perm{(34)}-\! 1 +\tfrac{5}{2} \perm{(24)}-\!\tfrac{3}{2} \perm{(14)})\diagKfive+ \nonumber\\
  & \phantom{\big[} \tfrac{1}{4} (1+ \perm{(34)} - \!2 \perm{(14)})\diagFfive+  \tfrac{1}{4} ( \perm{(34)}-\!1)\diagEfive + (\perm{(34)}-\!1)\diagQfive  + \nonumber\\
  & \phantom{\big[} \tfrac{1}{4} ( \perm{(24)}-1 )\diagLfive+ \tfrac{3}{2} ( 1+ \perm{(34)})\diagDfive  -\tfrac{1}{2} ( \perm{(34)}+ \perm{(14)})\diagOfive+ \nonumber\\
  & \phantom{\big[} \tfrac{1}{2} (\perm{(14)}-1)\diagMfive+ (\tfrac{1}{2} \perm{(34)}-\tfrac{3}{2} \perm{(14)})\diagPfive-\tfrac{1}{2} \diagNfive\big] F_{0} \Big\}
\end{align}
Interestingly the dimension of the kernel for $\eval{2221}$ is also $18$ and
by that not larger than the kernel for $\eval{2211}$. Naively one would expect
that the kernel dimension increases with growing Jordan-level $K:=\sum_i k_i$.
On the other hand the larger symmetry group ($S_3$ instead of $S_2 \times S_2$) reduces the kernel size which 
can even lead to a smaller kernel, as we will see in the case of $\eval{2222}$.
\begin{align}
  \mathrm{Ker}_{\eval{2221}} = & \mathcal{P}_{S_3} \Big\{ (2K_2-K_1) F_{1221}|^{d=1} + \nonumber\\
  & \big[ c K_1^2 + c' (K_2^2-K_1 K_2) \big] ( F_{1220}|^{d=2} + F_{2111} |^{d=2} + F_{0221} |^{d=2}) + \nonumber\\
  & K^{(3)} F_{0121}|^{d=4} + K_{S_4}^{(3)} F_{1111}|^{d=1} + \nonumber\\
  & \big[ c K_2(K_1 - K_2)(K_1-2K_2) + c' K_1^2 (K_1-2K_2) \big] F_{0220}|^{d=2}+\nonumber\\
  & \big[ c K_2(K_1-K_2)(K_1-2K_2) + c' K_1^2 (K_1-2K_2) \big] F_{2110}|^{d=2} + \nonumber\\
  & K_1^2 K_2 (K_1-K_2) F_{0111}|^{d=1}  + K_1^2 K_2 (K_1-K_2) F_{0120}|^{d=1} + \nonumber\\
  & K_1^2 K_2 (K_1-K_2) F_{1002}|^{d=1} \Big\}
\end{align}
There is not much surprise for most results. For logarithmic degree 1 and 2 we
get for the sets containing three $F$ a $d=1$ respectively a $d=2$ kernel. For degree 3 
we have the self-invariant term $F_{1111}$ with a $S_4$ symmetry and two $d=2$
kernels for $\{F_{0220}, F_{2020}, F_{2200}\}$ and $\{F_{2110}, F_{1210}, F_{1120}\}$.
There is also a set containing six $F$, which results in a full four dimensional $K^{(3)}$ kernel.

As expected combinatorial constraints show up the first time for degree four, because
of the last vertex having one leg only and thus disallowing any $l_{i4}^4$ ($i=1,2,3$) term.
For degree actually a $d=3$ kernel would have been possible, but eliminating all
$l_{i4}^4$ terms means that the kernel has to be reduced to a one-dimensional kernel each.
Also note that the only possible kernel term of degree 5 would have been $K_{S_4}^{(5)}$
which does not show up, because of the same combinatorial restriction.
\begin{align}
  \eval{2222} = &  \mathcal{P}_{S_4} \Big\{ \tfrac{1}{24}  F_{2222} + (\tfrac{1}{6} \perm{(13)}-\tfrac{1}{3})\diagAone F_{1222} + \nonumber\\
  & \big[ (\tfrac{1}{3} \perm{(12)}+\tfrac{1}{6} \perm{(14)})\diagBtwo-\tfrac{1}{3} \diagAtwo-\tfrac{1}{12} \perm{(13)}\diagCtwo \big] F_{0222} + \nonumber\\
  & \big[ \tfrac{1}{2} \perm{(24)}\diagAtwo+(\tfrac{1}{2} \perm{(23)}-\perm{(24)})\diagBtwo+\tfrac{1}{4} \perm{(13)(24)}\diagCtwo \big] F_{1122} + \nonumber\\
  & \big[ (3\perm{(34)}+3 +\perm{(14)})\diagAthree-5 \diagBthree - (\tfrac{13}{6}  +\tfrac{5}{2} \perm{(34)})\diagFthree +\nonumber\\
  & \phantom{\big[} (5 +3\perm{(24)})\diagCthree-  \diagDthree -(3 +\tfrac{3}{2} \perm{(14)})\diagEthree ) F_{1112} + \nonumber\\
  & \big[\tfrac{1}{2} (\perm{(124)}-11 -9 \perm{(12)}-7 \perm{(123)}- \perm{(132)}-3 \perm{(142)}-7 \perm{(14)}-7 \perm{(13)(24)}+ \nonumber\\
  & \phantom{\big[} \perm{(13)} -8\perm{(14)(23)})\diagAthree+(5 +\tfrac{3}{2} \perm{(24)}+3\perm{(14)}+\tfrac{9}{2} \perm{(13)(24)})\diagEthree+\nonumber\\
  & \phantom{\big[} (6 +7\perm{(23)}+5\perm{(12)})\diagBthree+(\tfrac{3}{2} +\perm{(13)}+\tfrac{5}{4} \perm{(13)(24)})\diagDthree+\nonumber\\
  & \phantom{\big[} (-10 -2\perm{(23)}-\perm{(24)}-8\perm{(12)}-5\perm{(123)}-2\perm{(132)})\diagCthree+\nonumber\\
  & \phantom{\big[} (9 +4\perm{(24)}+\tfrac{9}{2} \perm{(14)})\diagFthree \big] F_{0122}+ \nonumber\displaybreak[0]\\
  & \big[ \diagBfour -\tfrac{1}{3} \diagAfour-\tfrac{1}{2} \diagCfour- \diagDfour-\tfrac{1}{24} \diagEfour+\nonumber\\
  & \phantom{\big[} \tfrac{5}{12} \diagFfour+\tfrac{1}{6} \diagIfour+\tfrac{1}{3}  \diagKfour \big] F_{1111}+\displaybreak[0]\nonumber\\
  & \big[ \tfrac{1}{2}(\perm{(13)} - \perm{(23)})\diagAfour - \tfrac{1}{4} (1   + \perm{(13)(24)})\diagHfour+ (\tfrac{1}{8} \perm{(23)}-\!\tfrac{3}{8})\diagIfour+\nonumber\\
  & \phantom{\big[} \tfrac{1}{2} (1- \perm{(23)}+ \perm{(24)}-\tfrac{1}{2} \perm{(14)})\diagCfour - (\tfrac{1}{4}  +\tfrac{1}{8} \perm{(14)})\diagFfour -\tfrac{1}{2}  \diagKfour +\nonumber\\
  & \phantom{\big[} \tfrac{1}{2} (3+ \perm{(23)}- \perm{(24)} +3 \perm{(13)(24)})\diagDfour+  (\perm{(23)}-1-\perm{(1324)})\diagJfour +\nonumber\\
  & \phantom{\big[} \tfrac{1}{2} (\tfrac{1}{2} \perm{(24)} -1 - \perm{(23)}-\tfrac{1}{2} \perm{(13)})\diagBfour-\tfrac{1}{16} \diagEfour+\tfrac{1}{2}   \diagGfour\big] F_{0022}+\nonumber\displaybreak[0]\\
  & \big[ \tfrac{1}{4} \perm{(24)}\diagEfour + \tfrac{1}{2} (1 - \perm{(34)}+2\perm{(24)}+\tfrac{1}{2} \perm{(12)}-\tfrac{1}{2} \perm{(124)}+ \perm{(142)})\diagFfour+\nonumber\\ 
  & \phantom{\big[} \tfrac{1}{2}  (\perm{(134)}-\!2 \perm{(24)}-\!\perm{(13)}- \!\perm{(1243)}+\perm{(124)}-\!2 \perm{(142)}+ \perm{(143)}- \perm{(14)})\diagGfour+\nonumber\\
  & \phantom{\big[} \tfrac{1}{2} (3 \perm{(34)}\!-\!1 \!-\!3 \perm{(24)}\!-\! \perm{(12)(34)}\!+\!2\perm{(124)}\!+\!\perm{(12)}\!-\!2\perm{(13)(24)}\!-\!3 \perm{(142)})\diagDfour+ \nonumber\\
  & \phantom{\big[} (\perm{(24)} \!-\!\perm{(34)}\!- \!1 \!+\!\perm{(13)}\!-\!2\perm{(14)}\!-\!2\perm{(13)(24)})\diagJfour+(\perm{(34)}+2 )\diagKfour + \nonumber\\
  & \phantom{\big[} \tfrac{1}{2} (3 \perm{(34)}-\!1-\!\perm{(24)}+\perm{(12)}-\!2 \perm{(142)})\diagCfour+ \tfrac{1}{2} (\perm{(24)}- \perm{(13)})\diagHfour + \nonumber\\
  & \phantom{\big[} \tfrac{1}{2} ( \perm{(34)}-\perm{(12)}- \perm{(134)}-2\perm{(13)}+ \perm{(14)}+2 \perm{(143)})\diagAfour-\tfrac{1}{2} \diagIfour+ \nonumber\\
  & \phantom{\big[} \tfrac{1}{2} (1 -\perm{(34)}+3 \perm{(24)}+\perm{(13)}+3 \perm{(1243)}+\perm{(12)})\diagBfour  \big] F_{0112} + \nonumber\displaybreak[0]\\
  & \big[ \tfrac{1}{12} \perm{(13)}-\!\tfrac{4}{3} -\!\tfrac{1}{12} \perm{(14)})\diagGfive+(\tfrac{5}{12}  +\tfrac{11}{12} \perm{(13)})\diagHfive+\tfrac{4}{3} \perm{(14)}\diagJfive+\nonumber\\
  & \phantom{\big[} (\tfrac{13}{12} -\tfrac{1}{4} \perm{(14)})\diagAfive+ \tfrac{1}{12} (\perm{(13)}-\perm{(14)})\diagEfive+\tfrac{3}{4} (\perm{(12)}-1)\diagLfive + \nonumber\\
  & \phantom{\big[} (\tfrac{1}{6} \perm{(14)}-\tfrac{5}{6} \perm{(13)})\diagCfive+(\tfrac{1}{6} \perm{(13)}-\tfrac{7}{6} \perm{(14)})\diagDfive+\tfrac{1}{4} \perm{(14)}\diagNfive+\nonumber\\
  & \phantom{\big[} (\tfrac{1}{2}   -\tfrac{1}{6} \perm{(13)})\diagOfive+\tfrac{5}{6} (1 -\perm{(14)})\diagMfive- (\tfrac{5}{4}  +\tfrac{7}{12} \perm{(12)})\diagIfive+\nonumber\\
  & \phantom{\big[} (\tfrac{3}{2} \perm{(12)}-\tfrac{7}{6}  -\tfrac{2}{3} \perm{(14)})\diagKfive+\tfrac{2}{3} \diagBfive+  (\tfrac{1}{3}  -\tfrac{2}{3} \perm{(12)})\diagQfive + \nonumber\\
  & \phantom{\big[} (\tfrac{4}{3} \perm{(14)}-\tfrac{1}{6} -\tfrac{1}{2} \perm{(12)})\diagPfive + (\tfrac{1}{12} \perm{(14)}-\tfrac{5}{6}+\tfrac{3}{4} \perm{(13)})\diagFfive \big] F_{0111}+\nonumber\displaybreak[0]\\
  & \big[ \tfrac{1}{2}(1 - \perm{(243)} - \perm{(24)}-\tfrac{1}{2} \perm{(14)})\diagAfive+ \tfrac{1}{2} (1- \perm{(24)} + \perm{(23)}- \perm{(13)})\diagLfive + \nonumber\\
  & \phantom{\big[} \tfrac{1}{2} (\perm{(243)} -\!1+ \perm{(234)}+ \perm{(134)}-\! \perm{(143)})\diagGfive+( \perm{(243)}-\!\tfrac{1}{2}-\!\perm{(13)})\diagOfive +\nonumber\\  
  & \phantom{\big[} \tfrac{1}{2} (\perm{(234)}-\perm{(243)}+\perm{(13)}- \perm{(14)})\diagEfive+ (\perm{(23)}-\perm{(34)}-\tfrac{1}{2} \perm{(234)})\diagBfive+ \nonumber\\
  & \phantom{\big[} \tfrac{1}{2} ( \perm{(243)}- \perm{(24)}- \perm{(13)}+ \perm{(14)})\diagFfive+ (\perm{(34)}-\perm{(13)}+\perm{(13)(24)})\diagKfive+\nonumber\\
  & \phantom{\big[} (2\perm{(34)}+\perm{(234)}-\perm{(243)}+\perm{(13)}-\perm{(14)})\diagDfive-(1 +\tfrac{3}{4} \perm{(13)(24)})\diagHfive+\nonumber\\
  & \phantom{\big[} (3 +\tfrac{1}{4} \perm{(234)}-\perm{(243)}+\tfrac{1}{2} \perm{(13)})\diagIfive+-(\perm{(24)}+\perm{(1324)})\diagJfive+\nonumber\\
  & \phantom{\big[} \tfrac{1}{2} \perm{(1324)}\diagCfive - \tfrac{3}{8} \diagNfive -\perm{(13)}\diagPfive \big] F_{0012} + \nonumber\displaybreak[0]\\
  & \big[ \diagLsix + \diagWsix- \diagFsix+\tfrac{1}{4} \diagCsix+\tfrac{1}{2} \diagDsix-\tfrac{1}{4} \diagEsix + \nonumber\\
  & \phantom{\big[} \diagKsix  +\tfrac{1}{2}\diagBsix -\tfrac{1}{2}  \diagGsix-\tfrac{3}{2}  \diagHsix+\tfrac{1}{4}  \diagIsix - \diagMsix + \nonumber\\
  & \phantom{\big[} \tfrac{1}{4}  \diagJsix-5 \diagNsix-\tfrac{3}{4}  \diagOsix- \diagPsix-\tfrac{1}{2} \diagQsix- \diagRsix+\nonumber\\
  & \phantom{\big[} \tfrac{1}{2}  \diagTsix+3  \diagUsix+2 \diagVsix+\tfrac{5}{2}  \diagSsix-\tfrac{1}{6} \diagAsix \big] F_{0} \Big\}
\end{align}
We saw that the transition from $\eval{2211}$ to $\eval{2221}$ did not increase the 
dimension of the kernel mainly because of the increase of the discrete symmetry group
from $S_2 \times S_2$ to $S_3$. This transition to $\eval{2222}$ enlarges
the symmetry group from $S_3$ to $S_4$ and by that even reduces the dimension
of the kernel to $13$.
\begin{align}
  \mathrm{Ker}_{\eval{2222}} = & \mathcal{P}_{S_4} \Big\{ K_{S_2}^{(2)} F_{1122}|^{d=2} + \nonumber\\
  & \big[ c K_2(K_1-K_2)(K_1-2K_2) + c' K_1^2 (K1-2K_2)\big] F_{0122}|^{d=2}+ \nonumber\\
  & K_{S_4}^{(3)} F_{1112}|^{d=1} + K_{S_2}^{(4)} F_{0022}|^{d=3} + K_{S_4}^{(4)} F_{1111}|^{d=1} + \nonumber\\
  & \big[ c K_1^4 + c' K_2^2(K_1-K_2)^2 + c''(K_1^3K_2-2K_1 K_2^3+K_2^4) \big] F_{0112}|^{d=3} + \nonumber\\
  & K_1 K_2 (K_1-K_2)^2 (K_1+K_2) F_{0012}|^{d=1} \Big\}
\end{align}
There is no kernel of logarithmic degree one because at most we have
four $F$ in a set and the $S_4$ symmetry then is forbidden according to appendix \ref{overview of kernel}.
Additional combinatorial constraints start with logarithmic degree 5: no $l_{ij}^5$
for $ 1\le i<j\le 4$ is allowed to show up.

For $F_{0012}$ plus the five permutations there would have been a $d=3$ kernel,
but the given linear combination is the only one which eliminates all $l_{ij}^5$ terms.
For $\{F_{0111}, F_{1011}, F_{1101}, F_{1110}\}$ only $K_{S_4}^{(5,d=1)}$ 
would be possible, but is ruled by the combinatorial restriction.

That there is a kernel for $d=6$ is a bit unexpected. On the one hand we have (\ref{eq:s4 invariant kernel})
a two dimensional kernel, but on the other there are two constraints which need to be satisfied, namely
all $l_{ij}$ to the power of $5$ and to the power of $6$ have to be eliminated. The given combination
fulfills both restrictions and thus gives us an additional degree of freedom for $F_0$.

\section{Exact results for two logarithmic fields}\label{sec:exact results}
The most easiest non-trivial case is the one, where we have two logarithmic fields
and two primaries. For this case the correlator $\eval{k_1 k_2 0 0}$ for $k_1, k_2 > 0$ can be solved
exactly for arbitrary Jordan-rank $r$.

The correlator for Jordan-rank $r$ has the following form
\begin{align}
  \eval{k_1 k_2 0 0} = F_{k_1,k_2,0,0} + c_1 l_{12} F_{k_1-1,k_2,0,0} + c_2 l_{12} F_{k_1,k_2-1,0,0} + \ldots \; .
\end{align}
As described in subsection \ref{sec:restrictions} it is possible to identify some of 
the appearing $F$-terms with each other. In this case it turns out that it is easy to find
the identifications that stems from the integration process by inserting the above ansatz
in equation (\ref{gcwi_lcft1}). This leads to
\begin{align}
  O_1 \eval{k_1 k_2 0 0} & = -2 z_1 \eval{k_1\!-\!1, k_2, 0, 0} -2 z_2 \eval{k_1 , k_2\!-\!1,0,0} \; ,
\end{align}
and considering the terms of the lowest order in $\{l_{ij}\}$ only we get
\begin{align}  
  (z_1 + z_2) ( c_1 F_{k_1-1,k_2,0,0} + c_2 F_{k_1,k_2-1,0,0} ) + \mathcal{O}(l_{12}) \nonumber\\
  = - 2 z_1 F_{k_1-1,k_2,0,0} - 2 z_2 F_{k_1,k_2-1,0,0} + \mathcal{O}(l_{12}) \; . \label{eq:o1for2logs}
\end{align}
We immediately see that these equations do not have a solution.
As before we can circumvent the problem by reducing the complexity of
the equations, which can be accomplished by identification of some of
the functions $F$. Here we can solve equation (\ref{eq:o1for2logs}) by 
using the following identifications
\begin{align}
  F_{k_1-1,k_2,0,0} \equiv F_{k_1,k_2-1,0,0} \; .
\end{align}
This in perfect agreement with the results presented so far for $r=3$. 
Because of having the the above identifications we are left with only
one function $F$ for each logarithmic degree of $\eval{k_1 k_2 0 0}$.
Using (\ref{gcwi_lcft0}) yields after a short calculation the full result
for a correlator of Jordan-rank $r$ with two primary fields:
\begin{align}
   \eval{k_1 k_2 0 0} & = \sum_{n=0}^{k_1+k_2-(r-1)}  \frac{(-2)^n}{n!} l_{12}^n F_{k_1+k_2-(r-1)-n,r-1,0,0} \; . \label{eq:exact-result-for-2-logfields}
\end{align}
As a consistency check we can compare the above result with the
one presented in \cite{Flo01}, respectively \cite{RAK96}. 
For the two-point correlation function
the first paper gives the following result
\begin{align}
  \eval{\Psi_{k_1}(z_1) \Psi_{k_2}(z_2)} = \sum_{\ell=0}^{k_1+k_2} \frac{(-2)^\ell}{\ell!} l_{12}^\ell D_{(h_1=0,h_2=0,k_1+k_2-\ell)} \; . \label{eq:flo result}
\end{align}  
where we slightly adapted the notation and 
have set the conformal weights $h_1, h_2$ to zero.
The $D_{(\cdots)}$ are called ``structure constants" and have the property 
that $D_{(h,h;k)} = 0$ for $k< r-1$. In other words the index $\ell$ in (\ref{eq:flo result})
effectively runs from $0$ to $k_1+k_2 - (r-1)$ and thus 
(\ref{eq:exact-result-for-2-logfields}) and (\ref{eq:flo result})
are of identical structure. This means, that the polynomial dependence on 
the logarithms $l_{ij}$ is exactly the same and that precisely the same
number of free structure constants $D_{(\cdots)}$ or structure
functions $F_{\cdots}(x)$ are needed.

\section{Summary and discussion}
In the scope of this paper we analyzed the influence of the global conformal
symmetries in form of the global conformal Ward identities on 4-point
correlation functions in arbitrary logarithmic conformal field theory.
While it is not possible to completely determine the correlators, this 
does not even work in the CFT case, it is possible to fix the generic structure of the
correlators. 

The presented algorithm can be used to calculate the generic structure
of 4-point correlators. Within this paper we restricted ourselves to 
combinations of proper primary and logarithmic fields, but did
mention how to adjust the algorithm in order to extend the 
algorithm to pre-logarithmic fields. 

We explicitly gave the results for, up to permutations, all correlators
of Jordan-rank $r=2, 3$. In some of the results we found additional
constants which were identified as elements of the kernel $O$.
Furthermore we discussed various restrictions which limit the number
of terms that can appear in an ansatz or which lead to lesser degrees
of freedom in the kernel. Also we found that integration sometimes
requires that some functions $F$ need to be identified with each other.

Finally we gave explicit results for the case of exactly
two logarithmic fields for arbitrary Jordan-rank $r$. Studying
this very simple case showed us why we need to 
identify some of the functions $F$ with each other. Also we 
did a consistency check of the result and showed that 
equation (\ref{eq:exact-result-for-2-logfields}) is equivalent to 
the one presented in \cite{Flo01}. 

The comparison can be extended to three-point correlators.
For instance we can consider the terms of logarithmic degree $l=2$ of the 
correlator $\eval{2110}$ in a Jordan-rank $r=3$ theory,
cf. equation (\ref{eq:corr1120-r=3}):
\begin{align}
  \eval{2110}|_{l=2} = \big[ -\tfrac{1}{2} ( l_{12}^2 + l_{13}^2 +  l_{23}^2) +3l_{12}l_{13}+l_{12}l_{23}+l_{13}l_{23} \big]  F_{0} \; .
\end{align}
As a comparison we evaluate formula (3.11) in \cite{Flo01} and get for $l=2$ the same result,
\begin{align}
  \eval{211}|_{l=2} = \big[ -\tfrac{1}{2} ( l_{12}^2 + l_{13}^2 +  l_{23}^2) +3l_{12}l_{13}+l_{12}l_{23}+l_{13}l_{23} \big]  C_{(h_1,h_2,h_3;k=0)} \; ,
\end{align}
except that $F_0$ has to be replaced by the structure constant 
$C_{(h_1,h_2,h_3;k=0)}$. We once more note that we suppressed any direct but
trivial dependence on the conformal weights, so actually, we should compare
with $C_{(0,0,0;k=0)}$. However, our results are, up to the omitted 
prefactor $\prod_{i<j}z_{ij}^{\mu_{ij}}$, valid and independent of the
values of the conformal weights $h_i$.
For the other correlators like $\eval{2210}$ et cetera we also confirmed that the results
match if we restrict us to the highest logarithmic degree, which corresponds to $l^{\mathrm{max}}=k_1+k_2+k_3-r+1$.
As we will see in the following it is interesting to study the case where $l<l^{\mathrm{max}}$.
We use $\eval{2110}$ as an example again, but this time we consider the term of $l=1$ only:
\begin{align}
  \eval{2110}|_{l=1} = & F_{1020}(l_{13}-l_{12}-l_{23})+ F_{1110}(l_{23}-l_{12}-l_{13}) + \nonumber \\
                               & F_{1200}(l_{12}-l_{13}-l_{23}) \; .
\end{align}
We remind the reader that the above result includes the usual identifications
such as $F_{2100} \equiv F_{1200}$. The structure of the formula in \cite{Flo01}
makes it obvious that for $l=1$ only one structure constant shows up and thus the
corresponding term is
\begin{align}
  \eval{211}|_{l=1} =  -(l_{12}+l_{13}+l_{23} ) \, 
  C_{(h_1=0,h_2=0,h_3=0;k=1)} \; ,
\end{align}
where we again set the conformal weights to zero and slightly adjusted the
notation. Though looking differently at first glance we can achieve the same 
form of the result if we demand that the following extended identifications 
hold too, namely
\begin{align}
  F_{1200} \equiv F_{1020} \equiv F_{1110} \; .
\end{align}
This means that we do not only regain the $F_0$ terms,
but that we can reclaim all information, provided that we do 
all necessary identifications. With ``necessary'' we mean that
we have to identify all $F_{\cdots}$ terms of the same logarithmic
degree. 

We already encountered one situation where we had to
identify several functions $F$ with each other:
the initial conditions (\ref{eq:initcond1}) where we identified 
$F_0 \equiv F_0' \equiv \ldots$ by virtue of the cluster decomposition argument.

This evokes the question whether this form of massive identifications
of functions $F$ is necessary or useful in the context of some physical theory
respectively what conditions could force us to massively reduce
the number of functions $F$. It is clear that the special case where all
conformal weights $h_i$ are equal to each other has an additional symmetry,
since we can freely exchange the fields. In this case, we definitely expect
that a large number of such identifications should take place. 

Furthermore, one can quickly check that the given solutions remain
valid after identifying remaining free structure functions because any
remaining such function can be arbitrarily chosen as long as no further
constraints such as local conformal symmetry are invoked. Due to the
recursive dependence of the solutions for total Jordan-level $K$ on the ones
for level $K'<K$, identifications are consistent only if restricted to
functions $F_{k_1k_2k_3k_4}$, $F_{k_1'k_2'k_3'k_4'}$ with
$k_1+k_2+k_3+k_4=k_1'+k_2'+k_3'+k_4'$. However,
a more detailed analysis which identifications should be present in the
general case, i.\,e., for arbitrary values of the conformal weights $h_i$,
will be left to future work.

Of course, when all four fields in the 4-point function are logarithmic,
we cannot expect that the resulting polynomials in the $l_{ij}$ can be
matched with the ones of 2- and 3-point functions. But one might attempt
to make the following comparison. 

The structure functions 
$F_{k_1k_2k_3k_4}(x)$ are ultimately composed out of (a suitable 
generalization of) conformal blocks which depend on the internal
propagator in the 4-point function. Crossing symmetry of the 4-point function
imply that the structure functions possess for each asymptotic region
$|x|<1$, $|1-x|<1$, or $1/|x|<1$ expansions of the schematic form
\begin{align}
  F_{(h_1,k_1)(h_2,k_2)(h_3,k_3)(h_4,k_4)}(x) \sim
  \sum_{(h,k)} C_{(h_i,k_i)(h_j,k_j)}^{(h,k)}C^{ }_{(h,k)(h_l,k_l)(h_m,k_m)}
  + \ldots
\end{align}
for all permutations $\{i,j,l,m\}$ of $\{1,2,3,4\}$,
which must all be expansions of the same analytical functions. 
These expansions involve the 3-point
structure constants as well as the OPE structure constants. In the
logarithmic case, these structure ``constants'' are matrix valued with
coefficients in $\mathbb{C}[\{l_{ij}\}]$. In the notation used in this paper,
$C_{(h_1,k_1)(h_2,k_2)(h_3,k_3)} = \eval{k_1k_2k_3}$ where on both sides
all terms of the form $z_{ij}^{\mu_{ij}}$ depending in the canonical way
on the conformal weights are omitted. In the $r$-dimensional Jordan-cell
space, this defines matrices $(C_{k_1})_{k_2k_3}$ labeled by the first
Jordan-level and with indices given by the second and third Jordan-level.
In the same way, the propagator defines a matrix $(D)_{k_1k_2}=\eval{k_1k_2}$.
The OPE structure ``constants''
are then given by the matrix product 
\begin{align}
	(C_{k_1})_{k_2}^{\ k_3} = (C_{k_1})_{k_2 k}^{ }(D^{-1})^{k k_3}_{ }
\end{align}
involving the inverse propagator.
Now, one can compute the leading orders of the different expansions of the
4-point structure functions which will yield different polynomials
in the $l_{ij}$ with coefficients given by rational functions of the
2- and 3-point structure constants $D_{(h,h;p)}$ and $C_{(h_ih_j,h;q)}$.
Two observations can now be made: 

Firstly, the three expansions for the
$s$-, $t$- and $u$-channel, i.\,e., for $|x|<1$, $|1-x|<1$ and $1/|x|<1$ all
differ. They lead to different polynomials. It is easy to check in simple
examples that certain monomials in the $l_{ij}$ may appear only in one of
the expansions. This always happens for 4-point functions of the form
$\eval{k_1k_2k_3k_4}$ with all $k_i>0$ but not all $k_i$ equal.

Secondly, the polynomials in $l_{ij}$ with coefficients
given by the structure functions $F_{k_1k_2k_3k_4}(x)$ cannot be matched
to any of the three expansions. On the contrary, the 4-point functions
will involve all the different monomials in the $l_{ij}$ and in particular
all the ones which do not appear in all the expansions, but in only one of
them. It is therefore much more difficult to match the 4-point
structure functions to expressions in the 3- and 2-point structure
constants or to suggest further identifications as they can easily be
read off in the case of 4-point functions of type $\eval{k_1k_200}$ or
$\eval{k_1k_2k_30}$. In fact, it is not straightforward how the three
different expansions should be combined for a comparison of coefficients in
case all four fields are logarithmic. A further complication is given
by the freedom to change the polynomials in the $l_{ij}$ by elements in
the kernel of the operator $O$ or, equivalently, by a redefinition of the
structure function coefficients. But we believe that it would be very
interesting to investigate the consequences of crossing symmetry for the
structure functions of LCFT 4-point functions, because this might yield
severe restrictions on the number of functions which have to be determined
by other means, for example local conformal invariance. This is an
important task for future work in order to greatly ease the full computation of
4-point correlation functions in LCFT of rank $r>2$.

\subsubsection*{Acknowledgement}
The research of Michael Flohr is supported by the European Union
network HPRN-CT-2002-00325 and the research of Michael Flohr
and Marco Krohn is partially supported by the string theory network 
(SPP no. 1096), Fl 259/2-2, of the Deutsche Forschungsgemeinschaft.
M. Krohn would like to thank Sebastian Uhlmann and Robert Wimmer 
for helpful discussions.

\begin{appendix}

\section{Overview of the kernel terms}\label{overview of kernel}
The following tables contain the kernel terms that can show up for a logarithmic
degree from one to five. In addition to the logarithmic degree the kernel depends
on the number of functions $F$ that are involved and also on the discrete symmetry.
As we are considering four point functions only we are left with four different symmetry groups.

The format of the entries is the same as in subsection \ref{sec:additional constants}
with the small addition of the dimension $d$ of the kernel. It is interesting how similar the entries
for the different logarithmic degrees are, the only exception being the entry
for the pair $(F_3, S_3)$ respectively $(F_{12}, S_4)$. Also note that each column
contains the full kernel, namely if and only if $|F| = |S|$, where $S$ denotes the symmetry group
and $|S|$ its cardinality .

``$\leftrightarrow$" means that these entries have to be identical as
shown in (\ref{eq:s2 equivs s2xs2}), (\ref{eq:s3 equivs s4}).

\begin{center}
\begin{tabular}{c|cccccc}
 Log.deg 1 & $S_2$ & & $S_{2\times2}$ & $S_3$ &  & $S_4$ \\ 
\hline
 $F_{1}$  & $K_{S_2}^{(1),d=1}$ & $\leftrightarrow$ & $K_{S_2}^{(1),d=1}$  & 0 & $\leftrightarrow$ & 0  \\ 
 $F_{2}$  & $K^{(1),d=2}$ & & $K_{S_2}^{(1),d=1}$ & --- & & --- \\ 
 $F_{3}$  & --- & & --- & $(\ast)|^{d=1}$ &  & --- \\ 
 $F_{4}$  & --- & & $K^{(1),d=2}$ & --- & & 0  \\ 
 $F_{6}$  & --- & & --- & $K^{(1),d=2}$ & & $K_{S_2}^{(1),d=1}$\\ 
 $F_{12}$ & --- & & --- & --- &  &  $(\ast)|^{d=1}$ \\ 
 $F_{24}$ & --- & & --- & --- & & $K^{(1),d=2}$ \\ 
  \hline
  \multicolumn{7}{l}{$(\ast) = 2K_2 - K_1|^{d=1}$}
\end{tabular} 
\end{center}

\begin{center}
\begin{tabular}{c|cccccc}
 Log.deg 2 & $S_2$ & & $S_{2\times2}$ & $S_3$ &  & $S_4$ \\ 
\hline
 $F_{1}$  & $K_{S_2}^{(2),d=2}$ & $\leftrightarrow$ & $K_{S_2}^{(2),d=2}$  & $K_{S_4}^{(2),d=1}$ & $\leftrightarrow$ & $K_{S_4}^{(2),d=1}$  \\ 
 $F_{2}$  & $K^{(2),d=3}$ & & $K_{S_2}^{(2),d=2}$ & --- & & --- \\ 
 $F_{3}$  & --- & & --- & $(\ast)|^{d=2}$ &  & --- \\ 
 $F_{4}$  & --- & & $K^{(2),d=3}$ & --- & & $K_{S_4}^{(2),d=1}$ \\ 
 $F_{6}$  & --- & & --- & $K^{(2),d=3}$ & & $K_{S_2}^{(2),d=2}$  \\ 
 $F_{12}$ & --- & & --- & --- &  & $(\ast)|^{d=2}$ \\ 
 $F_{24}$ & --- & & --- & --- & & $K^{(2),d=3}$ \\ 
  \hline
  \multicolumn{7}{l}{$(\ast) = c K_1^2 + c' (K_2^2 - K_1 K_2)|^{d=2}$}
\end{tabular} 
\end{center}

\begin{center}
\begin{tabular}{c|cccccc}
 Log.deg 3 & $S_2$ & & $S_{2\times2}$ & $S_3$ &  & $S_4$ \\ 
\hline
 $F_{1}$  & $K_{S_2}^{(3),d=2}$ & $\leftrightarrow$ & $K_{S_2}^{(3),d=2}$  & $K_{S_4}^{(3),d=1}$ & $\leftrightarrow$ & $K_{S_4}^{(3),d=1}$  \\ 
 $F_{2}$  & $K^{(3),d=4}$ & & $K_{S_2}^{(3),d=2}$ & --- & & --- \\ 
 $F_{3}$  & --- & & --- &  $(\ast)|^{d=2}$ & & --- \\ 
 $F_{4}$  & --- & & $K^{(3),d=4}$ & --- &  & $K_{S_4}^{(3),d=1}$ \\ 
 $F_{6}$  & --- & & --- & $K^{(3),d=4}$ &  & $K_{S_2}^{(3),d=2}$ \\ 
 $F_{12}$ & --- & & --- & --- & & $(\ast)|^{d=2}$ \\ 
 $F_{24}$ & --- & & --- & --- & & $K^{(3),d=4}$ \\ 
  \hline
  \multicolumn{7}{l}{$(\ast) = c K_1^2(K_1-2K_2 ) + c' K_2(K_1-K_2)(K_1-2K_2)|^{d=2}$}\\
\end{tabular} 
\end{center}

\begin{center}
\begin{tabular}{c|cccccc}
 Log.deg 4 & $S_2$ & & $S_{2\times2}$ & $S_3$ &  & $S_4$ \\ 
\hline
 $F_{1}$  & $K_{S_2}^{(4),d=3}$ & $\leftrightarrow$ & $K_{S_2}^{(4),d=3}$  & $K_{S_4}^{(4),d=1}$ & $\leftrightarrow$ & $K_{S_4}^{(4),d=1}$  \\ 
 $F_{2}$  & $K^{(4),d=5}$ & & $K_{S_2}^{(4),d=3}$ & --- & & --- \\ 
 $F_{3}$  & --- & & --- &  $(\ast)|^{d=3}$ & & --- \\ 
 $F_{4}$  & --- & & $K^{(4),d=5}$ & --- &  & $K_{S_4}^{(4),d=1}$ \\ 
 $F_{6}$  & --- & & --- & $K^{(4),d=5}$ &  & $K_{S_2}^{(4),d=5}$ \\ 
 $F_{12}$ & --- & & --- & --- & & $(\ast)|^{d=3}$ \\ 
 $F_{24}$ & --- & & --- & --- & & $K^{(4),d=5}$ \\ 
  \hline
  \multicolumn{7}{l}{$(\ast) = c K_1^4 + c' K_2^2(K_1-K_2)^2 + c''(K_1^3 K_2 - 2 K_1 K_2^3 + K_2^4)$}
\end{tabular} 
\end{center}

\begin{center}
\begin{tabular}{c|cccccc}
 Log.deg 5 & $S_2$ & & $S_{2\times2}$ & $S_3$ &  & $S_4$ \\ 
\hline
 $F_{1}$  & $K_{S_2}^{(5),d=3}$ & $\leftrightarrow$ & $K_{S_2}^{(5),d=3}$  & $K_{S_4}^{(5),d=1}$ & $\leftrightarrow$ & $K_{S_4}^{(5),d=1}$  \\ 
 $F_{2}$  & $K^{(5),d=6}$ & & $K_{S_2}^{(5),d=3}$ & --- & & --- \\ 
 $F_{3}$  & --- & & --- &  $(\ast)|^{d=3}$ & & --- \\ 
 $F_{4}$  & --- & & $K^{(5),d=6}$ & --- &  & $K_{S_4}^{(5),d=1}$ \\ 
 $F_{6}$  & --- & & --- & $K^{(5),d=6}$ &  & $K_{S_2}^{(5),d=3}$ \\ 
 $F_{12}$ & --- & & --- & --- & & $(\ast)|^{d=3}$ \\ 
 $F_{24}$ & --- & & --- & --- & & $K^{(5),d=6}$ \\ 
  \hline
  \multicolumn{7}{l}{$(\ast) = c(-2K_1^3 K_2^2 + 8 K_1^2 K_2^3 - 11 K_1 K_2^4 + 5 K_2^5 ) + $}\\
  \multicolumn{7}{l}{$\phantom{(\ast) = }\; c'(-K_1^4 K_2 + 4K_1^3 K_2^2 -6 K_1^2 K_2^3 +4 K_1 K_2^4) +$} \\
  \multicolumn{7}{l}{$\phantom{(\ast) = }\; c''(8K_1^5 + 1 K_1^4 K_2  -10 K_1^2 K_2^3 +20 K_1 K_2^4 -20 K_2^5 )|^{d=3}$}
\end{tabular} 
\end{center}

\end{appendix}

\bibliographystyle{mk}
\bibliography{hep-th}

\end{document}